\documentclass[onecolumn,showpacs,11pt]{revtex4}
\usepackage{graphicx}
\usepackage{dcolumn}
\usepackage{bm}
\begin{document}
\newcommand{\hs}{\hspace*{0.5cm}}
\newcommand{\vs}{\vspace*{0.5cm}}
\newcommand{\be}{\begin{equation}}
\newcommand{\ee}{\end{equation}}
\newcommand{\bea}{\begin{eqnarray}}
\newcommand{\eea}{\end{eqnarray}}
\newcommand{\ben}{\begin{enumerate}}
\newcommand{\een}{\end{enumerate}}
\newcommand{\bde}{\begin{widetext}}
\newcommand{\ede}{\end{widetext}}
\newcommand{\nn}{\nonumber}
\newcommand{\crn}{\nonumber \\}
\newcommand{\Tr}{\mathrm{Tr}}
\newcommand{\non}{\nonumber}
\newcommand{\noi}{\noindent}
\newcommand{\al}{\alpha}
\newcommand{\la}{\lambda}
\newcommand{\bet}{\beta}
\newcommand{\ga}{\gamma}
\newcommand{\va}{\varphi}
\newcommand{\om}{\omega}
\newcommand{\pa}{\partial}
\newcommand{\+}{\dagger}
\newcommand{\fr}{\frac}
\newcommand{\bc}{\begin{center}}
\newcommand{\ec}{\end{center}}
\newcommand{\Ga}{\Gamma}
\newcommand{\de}{\delta}
\newcommand{\De}{\Delta}
\newcommand{\ep}{\epsilon}
\newcommand{\varep}{\varepsilon}
\newcommand{\ka}{\kappa}
\newcommand{\La}{\Lambda}
\newcommand{\si}{\sigma}
\newcommand{\Si}{\Sigma}
\newcommand{\ta}{\tau}
\newcommand{\up}{\upsilon}
\newcommand{\Up}{\Upsilon}
\newcommand{\ze}{\zeta}
\newcommand{\ps}{\psi}
\newcommand{\Ps}{\Psi}
\newcommand{\ph}{\phi}
\newcommand{\vph}{\varphi}
\newcommand{\Ph}{\Phi}
\newcommand{\Om}{\Omega}

\title{The $S_3$ flavor symmetry in 3-3-1 models}

\author{P. V. Dong}
\email {pvdong@iop.vast.ac.vn} \affiliation{Institute of Physics,
VAST, 10 Dao Tan, Ba Dinh, Hanoi, Vietnam}
\author{H. N. Long}
\email{hnlong@iop.vast.ac.vn} \affiliation{Institute of Physics,
VAST, 10 Dao Tan, Ba Dinh, Hanoi, Vietnam}
\author{C. H. Nam}
\affiliation{Institute of Physics, VAST, 10 Dao Tan, Ba Dinh,
Hanoi, Vietnam}
\author{V. V. Vien}
\affiliation{Department of Physics, Tay Nguyen University, 567 Le
Duan, Buon Ma Thuot, Vietnam}

\date{\today}

\begin{abstract}
We propose two 3-3-1 models (with either neutral fermions or
right-handed neutrinos) based on $S_3$ flavor symmetry responsible
for fermion masses and mixings. The models can be distinguished
upon the new charge embedding ($\mathcal{L}$) relevant to lepton
number. The neutrino small masses can be given via a cooperation
of type I and type II seesaw mechanisms. The latest data on
neutrino oscillation can be fitted provided that the flavor
symmetry is broken via two different directions $S_3\rightarrow
Z_2$ and $S_3\rightarrow Z_3$ (or equivalently in the sequel
$S_3\rightarrow Z_2 \rightarrow \{\mathrm{Identity}\}$), in which
the second direction is due to a scalar triplet and another
antisextet as small perturbation. In addition, breaking of either
lepton parity in the model with neutral fermions or lepton number
in the model with right-handed neutrinos must be happened due to
the $\mathcal{L}$-violating scalar potential. The TeV seesaw scale
can be naturally recognized in the former model. The degenerate
masses of fermion pairs ($\mu$, $\tau$), ($c,\ t$) and ($s,\ b$)
are respectively separated due to the $S_3\rightarrow Z_3$
breaking.

\end{abstract}


\maketitle

\section{\label{intro}Introduction}

The experiments of neutrino oscillations have indicated that the
neutrinos have small masses and mixings \cite{pdg}, thus the
standard model of fundamental particles and interactions must be
extended. Among the proposals known today for explanation of the
above problems, the seesaw mechanism~\cite{seesaw1} is perhaps the
most popular and natural. In this scenario, the heavy right-handed
neutrinos $\nu_R$ (or called neutral fermions $N_R$ in some
variants) are actually required so that the mechanism works. The
presence of these particles can imply interesting cosmological
consequences such as the baryon asymmetry via leptogenesis
\cite{leptog}. However, the mystery is that they have not been
observed. What is the natural origin of them. There have been nice
approaches in which they may be necessary constituents of the
theory such as left-right symmetry \cite{lrm} or $\mathrm{SO}(10)$
grand unification \cite{so10}.

An alternative is to extend the electroweak symmetry into
$\mathrm{SU}(3)_L \otimes \mathrm{U}(1)_X$, in which to complete
the fundamental representations of $\mathrm{SU}(3)_L$ with the
standard model doublets, the right-handed neutrinos or neutral
fermions may be acquired. This proposal has nice features and been
extensively studied over the last two decades, called 3-3-1 models
\cite{331m,331r,ecn331}. Indeed, in the standard model as well as
the theories mentioned the number of fermion families is left
arbitrarily although from the experimental observations and fits
we surely know that it is three. The reason possibly originates
from the fact that the anomalies are canceled on every family, no
interplay between families needed~\cite{anoma} (see also P. H.
Frampton in \cite{331m}). In the standard model this cancelation
is due to the chiral electroweak symmetry $\mathrm{SU}(2)_L\otimes
\mathrm{U}(1)_Y$ with the relevant $\mathrm{SU}(2)_L$ trace
$\mathrm{Tr}[\{T_a,T_b\}T_c]=0$ for any fermion representation.
The simplest extension to the $\mathrm{SU}(3)_L \otimes
\mathrm{U}(1)_X$ thus implies the trace nonvanished, thereby all
the families have to be taken into account which follows that the
number of fermionic triplets equals to those of antitriplets.
Consequently, the number of families is a integral multiple of the
fundamental color number, which is three, coinciding with the
observation.

There are two typical versions of the 3-3-1 models concerning
respective lepton sectors. In the first version, called minimal
3-3-1 model, three $\mathrm{SU}(3)_L$ lepton triplets take the
form $(\nu_L,l_L,l^c_R)$ in which $l_{R}$ are the ordinary
right-handed charged-leptons \cite{331m}. In the second version,
the third components of lepton triplets respectively include
right-handed neutrinos, $(\nu_L,l_L,\nu^c_R)$, called 3-3-1 model
with right-handed neutrinos \cite{331r}. In Refs.
\cite{dlshA4,dlsvS4}, we have proposed another variant of the
lepton sectors as $(\nu_L,l_L,N^c_R)$ where $N_R$ are three new
fermion singlets carrying no lepton-number in contradiction to
that of the right-handed neutrinos, called 3-3-1 model with
neutral fermions. Among the 3-3-1 models as mentioned, the last
one can recover the tribimaximal neutrino-mixing form \cite{hps}
under $A_4$ and $S_4$ flavor symmetries, respectively. For some
other examples based on these flavor symmetries, see \cite{A4} and
\cite{S4}. We notice also that in the minimal 3-3-1 model the
charged-lepton masses can be naturally generated via the
contribution of $\mathrm{SU}(3)_L$ scalar antisextet. In the two
others the neutrino masses by contrast can be arisen similarly
\cite{dlprd,dlshA4,dlsvS4}.

The parameters of neutrino oscillations such as the squared mass
differences and mixing angles are now very constrained. The data
in PDG2010 \cite{pdg} imply \bea s^2_{23}&=&0.5,\hs
s^2_{12}=0.304,\hs s^2_{13}<0.035,\crn \Delta
m^2_{21}&=&7.65\times 10^{-5}\ \mathrm{eV}^2,\hs |\Delta
m^2_{31}|=2.40\times 10^{-3}\ \mathrm{eV}^2,\label{olddata}\eea
where (and hereafter) the best fits are taken into accounts.
Whereas, under the light of new experiments \cite{mal}, the new
data \cite{vale} (see also \cite{flivale} for previous fits) have
been given to be slightly modified from the old fits
(\ref{olddata}):\bea s^2_{23}&=&0.52,\hs s^2_{12}=0.312,\hs
s^2_{13}=0.013,\crn \Delta m^2_{21}&=&7.59\times 10^{-5}\
\mathrm{eV}^2,\hs |\Delta m^2_{31}|=2.50\times 10^{-3}\
\mathrm{eV}^2.\label{newdata}\eea If such conclusions on the
mixing angles are exact, the simplest explanation is probably due
to a $S_3$ flavor symmetry which is the smallest non-Abelian
discrete group \cite{kj}. In fact, there is an approximately
maximal mixing of two flavors $\mu$ and $\tau$ as given above
which can be connected by a $\underline{2}$ irreducible
representation of $S_3$. Besides the $\underline{2}$, the group
$S_3$ can provide two inequivalent singlet representations
$\underline{1}$ and $\underline{1}'$ which play a crucial role in
reproducing consistent fermion masses and mixings. The $S_3$
models have been studied extensively over the last decade
\cite{s3model}.

We would like to extend the above application to the two latter
3-3-1 models with respect to the right-handed neutrinos $\nu_R$
and the neutral fermions $N_R$ as mentioned because of the
following independent issues. (i) The observed neutrino masses can
be obtained by the seesaw mechanism; (ii) The anomaly cancelation
as determined requires also that one family of quarks has to
transform under $\mathrm{SU}(3)_L$ differently from the two
others. We should therefore search for a family symmetry with
$\underline{2}$ and $\underline{3}$ representations respectively
acting on such 2- and 3-family indices. Looking for a group with
corresponding irreducible representations, the simplest is $S_4$
which has been explored in \cite{dlsvS4}. Another possibility with
$A_4$ has been given in \cite{dlshA4}. In this paper, it is worth
to investigate a simpler group choice with $S_3$ in which
$\underline{3}$ and $\underline{2}$ are the defining and
irreducible representations of this group, respectively. The
physics as we will see is different from the formers
\cite{dlshA4,dlsvS4}. It is also noted that the numbers of fermion
families in the 3-3-1 model have an origin from the anomaly-free
gauge symmetry naturally meet with our criteria on the dimensions
of flavor group representations as such $S_3$, unlike the others
in the literature put by hand \cite{s3model,A4,S4}.

The rest of this work is as follows. In Sec. \ref{model} we
present the necessary elements of the 3-3-1 model with neutral
fermions $N_R$ under the $S_3$ symmetry as well as introducing
necessary Higgs fields responsible for the quark and
charged-lepton masses. Section \ref{neutrinomass} is devoted to
the neutrino mass problem. Section \ref{331RH} introduces the
$S_3$ symmetry into the 3-3-1 model with right-handed neutrinos
and briefly remarks on its consequences. We summarize our results
and make conclusions in Sec.~\ref{conclus}. Appendix \ref{apa}
briefly provides the theory of $S_3$ group. Appendices \ref{apb}
and \ref{apc} present the lepton numbers and scalar potentials of
both the models, respectively.

\section{\label{model}The 3-3-1 model with neutral fermions ($N_R$)}

\subsection{Fermion content}

The gauge symmetry is given by $\mathrm{SU}(3)_C\otimes
\mathrm{SU}(3)_L \otimes \mathrm{U}(1)_X$ (thus named 3-3-1),
where the electroweak factor $\mathrm{SU}(3)_L \otimes
\mathrm{U}(1)_X$ is extended from those of the standard model
whereas the strong interaction sector is retained. Each lepton
family including a new electrically- and leptonically-neutral
chiral fermion $(N_R)$ is arranged under the $\mathrm{SU}(3)_L$
symmetry as a triplet $(\nu_L, l_L, N^c_R)$ and a singlet $l_R$.
The residual electric charge operator $Q$ is related to the
generators of the gauge symmetry by $Q=T_3-\fr{1}{\sqrt{3}}T_8+X$,
where $T_a$ $(a=1,2,...,8)$ are $\mathrm{SU}(3)_L$ charges with
$\mathrm{Tr}T_aT_b=\fr 1 2 \de_{ab}$ and $X$ is the charge of
$\mathrm{U}(1)_X$. This means that the model considered does not
contain exotic electric charges in the fundamental fermion, scalar
and adjoint gauge-boson representations.

The lepton number is also a residual charge and not commuting with
the gauge symmetry unlike the standard model. It is better to work
with a new conserved charge $\mathcal{L}$ commuting with the gauge
symmetry and related to the ordinary lepton number by diagonal
matrices $L=\fr{2}{\sqrt{3}}T_8+\mathcal{L}$
\cite{clong,dlshA4,dlsvS4}. This is only convenient for accounting
the global lepton numbers of model particles since $T_8$ is a
gauged charge, consequently $L$ is gauged which contrasts with
outset. The $T_8$ can be understood as the charge of a group
replication of $\mathrm{SU}(3)_L$ but globally taken. The lepton
charge arranged in this way (i.e. $L(N_R)=0$ as assumed) will
prevent unwanted interactions (due to $\mathrm{U}(1)_\mathcal{L}$
symmetry) and symmetry-breakings (due to the lepton parity as
shown below) providing consistent lepton and quark spectra with
distinguish phenomena from those in the 3-3-1 model with
right-handed neutrinos as presented in Sec. \ref{331RH}. By this
embedding, the model also does not contain exotic leptonic charges
in the fundamental fermion, scalar and adjoint gauge-boson
representations, e.g. exotic quarks $U,D$ as well as new
non-Hermitian gauge bosons $X^0$, $Y^\pm$ possess lepton charges
as of the ordinary leptons: $L(D)=-L(U)=L(X^0)=L(Y^{-})=1$.

A brief of the theory of $S_3$ group is given in Appendix
\ref{apa}. The $S_3$ contains one doublet irreducible
representation $\underline{2}$ and two singlets $\underline{1}$,
$\underline{1}'$. As motivated by assigning the flavor
$\underline{2}$ and $\underline{3}$ contents in which the
$\underline{3}$ in this case is defining representation decomposed
as $\underline{3}=\underline{2}\oplus \underline{1}$, we should
therefore put all the model fermions in $\underline{1}$ and
$\underline{2}$. To be concrete, we put the first family fermions
in $\underline{1}$, while the two other families are in
$\underline{2}$. Under the $[\mathrm{SU}(3)_L, \mathrm{U}(1)_X,
\mathrm{U}(1)_\mathcal{L},\underline{S}_3]$ symmetries as proposed
the fermions correspondingly transform as follows \bea \psi_{1L}
&=& \left(
    \nu_{1L},\ l_{1L},\ N^c_{1R}\right)^T\sim
    [3,-1/3,2/3,\underline{1}],\hs
    l_{1R}\sim[1,-1,1,\underline{1}],
    \crn \psi_{ \alpha L }&=&
\left(\nu_{\alpha L},\ l_{\alpha L},\ N^c_{\alpha R} \right)^T
\sim [3,-1/3,2/3,\underline{2}],\hs l_{\alpha
R}\sim[1,-1,1,\underline{2}], \crn
 Q_{1L}&=& \left(u_{1L},\ d_{1L},\
 U_{L}\right)^T\sim[3,1/3,-1/3,\underline{1}],\crn
u_{1R} &\sim &[1,2/3,0,\underline{1}],\hs
d_{1R}\sim[1,-1/3,0,\underline{1}],\hs U_R\sim
[1,2/3,-1,\underline{1}],\label{conts3}\\
 Q_{\al L}&=&\left(d_{\alpha L},\ -u_{\alpha L},\ D_{\alpha
 L}\right)^T\sim[3^*,0,1/3,\underline{2}],\crn
u_{\al R}&\sim& [1,2/3,0,\underline{2}],\hs d_{\al
R}\sim[1,-1/3,0,\underline{2}],\hs D_{\al R}
\sim[1,-1/3,1,\underline{2}].\nn\eea where $\al=2,3$ is a family
index of the last two lepton and quark families, which are in
order defined as the components of the $\underline{2}$
representations. Note that the $\underline{2}$ for quarks meets
the requirement of anomaly cancelation where the last two
left-quark families are in $3^*$ while the first one as well as
the leptons are in $3$. All the $\mathcal{L}$ charges of the model
multiplets are listed in the square brackets.

In the following, we consider possibilities for generating the
fermion masses. The scalar multiplets needed for this purpose
would be introduced accordingly.

\subsection{Charged-lepton mass}

To generate masses for the charged leptons, we need two scalar
multiplets:
\bea \phi = \left(%
\begin{array}{c}
  \phi^+_1 \\
  \phi^0_2 \\
  \phi^+_3 \\
\end{array}%
\right)\sim [3,2/3,-1/3, \underline{1}],\hs \phi' = \left(%
\begin{array}{c}
  \phi'^+_1 \\
  \phi'^0_2 \\
  \phi'^+_3 \\
\end{array}%
\right)\sim [3,2/3,-1/3, \underline{1}'] \eea with VEVs $\langle
\phi \rangle = (0,v,0)^T$ and $\langle \phi' \rangle =
(0,v',0)^T$. Notice that the numbered subscripts are the indices
of $\mathrm{SU}(3)_L$. The VEV of $\phi$ conserves $S_3$ while
that of $\phi'$ breaks this symmetry. Here the three elements of
$S_3$ corresponding to interchanges of two within three objects
are broken. Therefore the $S_3$ breaking in the charged lepton
sector is $S_3\rightarrow Z_3$ which consists of the identity
element and the two total permutations.

The Yukawa interactions are \bea -\mathcal{L}_{l}=h_1
\bar{\psi}_{1L} \phi l_{1R}+h (\bar{\psi}_{2 L} l_{2
R}+\bar{\psi}_{3 L} l_{3 R})\phi +h' (\bar{\psi}_{3L}
l_{3R}-\bar{\psi}_{2 L} l_{2 R}) \phi' +h.c. \label{Ycpl1}\eea The
mass Lagrangian reads $$
-\mathcal{L}^{\mathrm{mass}}_l=(\bar{l}_{1L},\bar{l}_{2L},\bar{l}_{3L})
M_l (l_{1R},l_{2R},l_{3R})^T+ h. c.$$  where
 \bea M_l=
\left(%
\begin{array}{ccc}
  h_1v & 0 & 0 \\
   0 &  h v-h' v'& 0\\
  0 &   & h v+h' v'\\
\end{array}%
\right)\equiv\left(%
\begin{array}{ccc}
  m_e & 0 & 0 \\
   0 &  m_\mu & 0\\
  0 &   & m_\tau\\
\end{array}%
\right),\eea which has the diagonal form. The diagonalization
matrices are therefore $U_{lL}= U_{lR}=1.$ This means that the
charged leptons $l_{1,2,3}$ by themselves are the physical mass
eigenstates. The lepton mixing matrix depends on only that of the
neutrinos that will be studied in the next section. The masses of
muon and tau are explicitly separated by $\phi'$ resulting from
the breaking $S_3\rightarrow Z_3$. This is why we introduce
$\phi'$ in accompanying with $\phi$.

The experimental mass values for the charged leptons at the weak
scale are given as \cite{pdg}: \bea m_e=0.511\ \textrm{MeV},\hs \
m_{\mu}=106.0 \ \textrm{MeV},\hs m_{\tau}=1.77\ \textrm{GeV} \eea
Thus, we get \bea h_1 v=0.511\ \textrm{MeV},\ hv =938\
\textrm{MeV},\   h^\prime v'=832 \ \textrm{MeV} \eea It follows
that $h_1\ll h \sim h'$, provided $v'\sim v$. We particularly
notice that the $\mu-\tau$ mass splitting term due to the
$S_3\rightarrow Z_3$ breaking is necessarily large like that of
the $S_3$ conserving, which are all given in the scale of one haft
tau mass.

\subsection{\label{quark}Quark mass}

To generate the quark masses, we additionally introduce the
following scalar multiplets: \bea \chi&=& \left(\chi^0_1,  \
  \chi^-_2, \  \chi^0_3 \right)^T\sim[3,-1/3,2/3,\underline{1}],\crn
   \eta&=&
\left(\eta^0_1, \  \eta^-_2, \
\eta^0_3\right)^T\sim[3,-1/3,-1/3,\underline{1}],\\ \eta' &=&
\left(
  \eta'^0_1, \  \eta'^-_2, \  \eta'^0_3\right)^T
  \sim[3,-1/3,-1/3,\underline{1}'].\nn\eea It is noticed that
these scalars do not couple to the lepton sector due to the gauge
invariance. The Yukawa interactions are then \bea -\mathcal{L}_q
&=& f_1 \bar{Q}_{1L}\chi U_R + f \bar{Q}_{L}\chi^* D_{R}\crn &&
+h^u_{1} \bar{Q}_{1L}\eta u_{1R}  + h^d \bar{Q}_{ L}\eta^*
d_{R}+h'^d \bar{Q}_{ L}\eta'^* d_{R}\crn && +h^d_{1}
\bar{Q}_{1L}\phi d_{1R}+ h^u \bar{Q}_{ L}\phi^* u_{R}+ h'^u
\bar{Q}_{ L}\phi'^* u_R\crn &&+h.c.\eea

We now introduce a residual symmetry of lepton number $P_l\equiv
(-1)^L$, called ``lepton parity'' \cite{dlshA4,lpa}, in order to
suppress the mixing between ordinary quarks and exotic quarks. For
a summary of lepton number of the model particles, see Appendix
\ref{apa}. The particles with even parity ($P_l=1$) have $L=0,\pm
2$ such as $N_R$, ordinary quarks and gauge bosons, the new
neutral gauge boson $Z'$, $\phi_{1,2}$, $\phi'_{1,2}$,
$\eta_{1,2}$, $\eta'_{1,2}$, $\chi_3$, and so on. The odd parity
particles $(P_l=-1)$ possess $L=\pm 1$ such as ordinary leptons,
$U$, $D$, the new non-Hermitian $X$ and $Y$, $\phi_3$, $\phi'_3$,
$\eta_3$, $\eta'_3$, $\chi_{1,2}$, and so forth. In this framework
we assume that the lepton parity is an exact symmetry, not
spontaneously broken. This means that $\eta_3$, $\eta'_3$ and
$\chi_1$ cannot develop VEV, and the concerning phenomena will be
skip. However, a brief discussion of broken lepton parity for the
quark sector is given at the end. For the neutrino sector we
always suppose, however, that the lepton parity is broken so that
the neutrino mass matrix taken into account is the most general.
Otherwise, the corresponding VEVs that carry odd parity will
vanish. The general conclusions obtained for the lepton sector are
unchanged because the lepton parity commutes with $S_3$.

Suppose that the VEVs of $\eta$, $\eta'$ and $\chi$ are $u$, $u'$
and $w$, where $u=\langle \eta^0_1\rangle$, $u'=\langle
\eta'^0_1\rangle$, $w=\langle \chi^0_3\rangle$ (the other VEVs
$\langle \eta^0_3\rangle$, $\langle \eta'^0_3\rangle$, and
$\langle\chi^0_1\rangle$ vanish due to the lepton parity
conservation). The exotic quarks therefore get masses $m_U=f_1 w$
and $m_{D_{1,2}}=f w$. In addition, $w$ has to be much larger than
those of $\phi$, $\phi'$, $\eta$ and $\eta'$ for a consistency
with the effective theory. The mass matrices for ordinary
up-quarks and down-quarks are, respectively, obtained as follows:
\bea M_u&=&
\left(%
\begin{array}{ccc}
  h^u_1 u & 0 & 0 \\
  0 & h^u v+h^{\prime u} v' & 0 \\
  0 & 0 & h^u v - h^{\prime u} v' \\
\end{array}%
\right)\equiv\left(%
\begin{array}{ccc}
  m_u & 0 & 0 \\
  0 & m_c & 0 \\
  0 & 0 & m_t \\
\end{array}%
\right), \crn M_d&=&
\left(%
\begin{array}{ccc}
   h^d_1 v & 0 & 0 \\
  0 &h^d u+h^{\prime d} u' & 0 \\
  0 & 0 & h^d u - h^{\prime d} u' \\
\end{array}%
\right)
\equiv\left(%
\begin{array}{ccc}
  m_d & 0 & 0 \\
  0 & m_s & 0 \\
  0 & 0 & m_b \\
\end{array}%
\right).\label{qmass}\eea In similarity to the charged lepton
sector, the masses of $c-t$ and $s-b$ quarks are (in pair)
separated by the scalars $\phi'$ and $\eta'$ due to the
$S_3\rightarrow Z_3$ symmetry breaking, respectively. We see that
the introduction of $\eta'$ is necessary to provide the different
masses of $s$ and $b$ quarks. The current mass values for the
quarks are given by \cite{pdg} \bea m_u&=&(1.5\div3.3)\
\textrm{MeV},\hs  m_d=(3.5\div 6.0)\ \textrm{MeV},\hs
m_c=(1.16\div1.34)\  \textrm{GeV},\crn m_s&=&(70.0\div130.0)\
\textrm{MeV},\hs  m_t=(169.0\div 173.3)\ \textrm{GeV},\hs
m_b=(4.13\div4.37)\ \textrm{GeV}.\label{vien3}\eea Therefore we
have \bea h^u_{1}u & = & (1.5 \div 3.3 ) \ \textrm{MeV},\,
h^d_{1}v = (3.5 \div 6.0) \ \textrm{MeV} \, , h^uv= (85.08 \div
87.32) \ \textrm{GeV},\crn h^du & = & (2.10 \div 2.25) \
\textrm{GeV},\, h'^uv' = - (83.83 \div 86.07) \ \textrm{GeV}\, ,
h'^d u'= - (2.00 \div 2.15) \ \textrm{GeV}.\label{huyen3}\eea  If
$u\sim v \sim u' \sim v'$, the Yukawa coupling hierarchies are
$h^u_1,\ h^d_1\ll h^d,\ |h'^d| \ll h^u,\ |h'^u|$. It is to be
noted that the $S_3$ breaking terms in this case are also large in
comparison to the conserving ones.

The unitary matrices which couple the left-handed quarks $u_L$ and
$d_L$ to those in the mass bases are $U_{uL}=1$ and $U_{dL}=1$,
respectively. The CKM quark mixing matrix at the tree level is
then \bea U_\mathrm{CKM}=U^{\dagger}_{dL} U_{uL}=1.\label{a41}\eea
This is a good approximation for the realistic quark mixing
matrix, which implies that the mixings among the quarks are
dynamically small. The small permutations such as a breaking of
the lepton parity due to the odd VEVs $\langle \eta^0_3\rangle$,
$\langle \eta'^0_3\rangle$, $\langle\chi^0_1\rangle$, or a
violation of $\mathcal{L}$ and/nor $S_3$ symmetry due to unnormal
Yukawa interactions, namely $\bar{Q}_{1L}\chi u_{1R}$, $\bar{Q}_L
\chi^* d_R$, $\bar{Q}_{1L}\chi u_{R}$ and so forth, will disturb
the tree level matrix resulting in mixing between ordinary and
exotic quarks and possibly providing the desirable quark mixing
pattern \cite{dlsvS4,dlshA4}. This also leads to the flavor
changing neutral current at the tree level but strongly suppressed
\cite{dlsvS4,dlshA4}. See also Section \ref{331RH} for a similar
matter encountered in the 3-3-1 model with right-handed neutrinos.
A detailed study on these matters are out of the scope of this
work and should be skep.

\section{\label{neutrinomass}Neutrino masses and mixing}

The neutrino masses arise from the couplings of $\bar{\psi}^c_L
\psi_L$ to scalars, where $\bar{\psi}^c_L \psi_L$ transforms as
$3^*\oplus 6$ under $\mathrm{SU}(3)_L$. Notice that in the first
term of decomposition the $\psi_{1,2,3}$ are totally antisymmetric
in flavor indices, while they are symmetric in the second term.
For the known scalar triplets, only available interactions are
$(\bar{\psi}^c_{ 2L} \psi_{3 L}-\bar{\psi}^c_{3L} \psi_{2
L})\phi'$, but explicitly suppressed because of the
$\mathcal{L}$--symmetry. We will therefore propose a new SU(3)$_L$
antisextet instead coupling to $\bar{\psi}^c_{ L}\psi_{ L}$
responsible for the neutrino masses. The antisextet transforms as
\bea s_i =
\left(%
\begin{array}{ccc}
  s^0_{11} & s^+_{12} & s^0_{13} \\
  s^+_{12} & s^{++}_{22} & s^+_{23} \\
  s^0_{13} & s^+_{23} & s^0_{33} \\
\end{array}%
\right)_i \sim [6^*,2/3,-4/3,\underline{2}], \eea where the
numbered subscripts on the component scalars are the
$\mathrm{SU}(3)_L$ indices, whereas $i=1,2$ is that of $S_3$. Note
that $i$ and $\al$ as mentioned belong to the same index kind. The
VEV of $s$ is set as $(\langle s_1\rangle,\langle s_2\rangle)$
under $S_3$, in which \bea
\langle s_i\rangle=\left(%
\begin{array}{ccc}
  \la_{i } & 0 & v_{i} \\
  0 & 0 & 0 \\
  v_{i} & 0 & \Lambda_{i } \\
\end{array}%
\right). \label{s1}\eea

Following the potential minimization conditions, we have several
VEV aligments. The first one is that $\langle s_1\rangle=\langle
s_2\rangle$ then $S_3$ is broken into $Z_2$ consisting of the
identity element and one interchange (within the three) of $S_3$.
The second one is that $\langle s_1\rangle\neq 0=\langle
s_2\rangle$ or $\langle s_1\rangle=0\neq \langle s_2\rangle$ then
$S_3$ is broken into $Z_3$ like the case of the charged lepton
sector. To obtain a realistic neutrino spectrum, in this work we
argue that both the breakings $S_3\rightarrow Z_2$ and
$S_3\rightarrow Z_3$ must be taken place. However, the VEVs of $s$
does only one of these tasks. We therefore assume that its VEVs
are aligned as the former to derive the first direction of the
breakings $S_3\rightarrow Z_2$, and this happens in any case
below: \be \la_1=\la_2\equiv \la_s,\hs v_1=v_2\equiv v_s,\hs
\La_1=\La_2\equiv \La_s.\ee

To achieve the second direction of the breakings $S_3\rightarrow
Z_3$, we additionally introduce another scalar which lies in
either $\underline{1}'$ or $\underline{2}$ (with the second
alignment of VEVs as mentioned above). However, this scalar is
also equivalent to break the $Z_2$ subgroup of the first
direction. We can therefore understand the misalignment of the
VEVs as follows. The $S_3$ is broken via two stages, the first
stage is $S_3\rightarrow Z_2$ and the second stage is
$Z_2\rightarrow \{\mathrm{identity}\}$ (instead of $S_3\rightarrow
Z_3$). The second stage (or direction) can be achieved within each
case below. \ben
\item A new $\mathrm{SU}(3)_L$ triplet $\rho$ (if the
$\mathcal{L}$--symmetry is allowed), which is impossible in
$\underline{2}$ since $(\bar{\psi}^c_L \psi_L)_{\underline{2}}
=(\bar{\psi}^c_{3L} \psi_{3 L},\bar{\psi}^c_{2L} \psi_{2 L}$)=0
due to antisymmetric in $\psi_{2}$ and  $\psi_{3}$, is thus put in the $\underline{1}'$: \be \rho = \left(%
\begin{array}{c}
  \rho^+_1 \\
  \rho^0_2 \\
  \rho^+_3 \\
\end{array}%
\right)\sim [3,2/3,-4/3, \underline{1}'], \ee with the VEVs given
by $\langle \rho \rangle = (0,v_\rho,0)^T$. \item Another
antisextet $s'$, which is impossible in $\underline{1}'$ since
$(\bar{\psi}^c_L \psi_L)_{\underline{1}'} =\bar{\psi}^c_{ 2L}
\psi_{3 L}-\bar{\psi}^c_{3L} \psi_{2 L}=0$ due to symmetric in
$\psi_2$ and $\psi_3$, is thus left with the $\underline{2}$ with
VEVs chosen by \bea s'_i =
\left(%
\begin{array}{ccc}
  s'^0_{11} & s'^+_{12} & s'^0_{13} \\
  s'^+_{12} & s'^{++}_{22} & s'^+_{23} \\
  s'^0_{13} & s'^+_{23} & s'^0_{33} \\
\end{array}%
\right)_i \sim [6^*,2/3,-4/3,\underline{2}],\hs \langle s'_1\rangle=\left(%
\begin{array}{ccc}
  \la'_{s} & 0 & v'_{s} \\
  0 & 0 & 0 \\
  v'_{s} & 0 & \Lambda'_{s} \\
\end{array}%
\right),\hs \langle s'_2\rangle=0. \label{s1} \eea \een

In calculation, combining both cases we have the Yukawa
interactions:
 \bea -\mathcal{L}_\nu&=& \fr 1 2 x
(\bar{\psi}^c_{2 L} \psi_{2 L} s_{1}+\bar{\psi}^c_{3 L} \psi_{3 L}
s_{2})+\fr 1 2 y \bar{\psi}^c_{1 L} (\psi_{2 L} s_{2 }+\psi_{3 L}
s_{1 })\crn &&+\fr 1 2 x' (\bar{\psi}^c_{2 L} \psi_{2 L}
s'_{1}+\bar{\psi}^c_{3 L} \psi_{3 L} s'_{2})+\fr 1 2 y'
\bar{\psi}^c_{1 L} (\psi_{2 L} s'_{2 }+\psi_{3 L} s'_{1 })\crn
&&+\fr 1 2 z(\bar{\psi}^c_{ 2L} \psi_{3 L}-\bar{\psi}^c_{3L}
\psi_{2 L})\rho \crn &&+h.c., \label{s3nm1} \eea where the
couplings $y$, $y'$ and $z$ are of lepton flavor changing
interactions. The mass Lagrangian for the neutrinos is given by
\bea -\mathcal{L}^{\mathrm{mass}}_\nu &=&\fr 1 2 x
(\la_{s}\bar{\nu}^c_{2 L}\nu_{2L}+v_{s}\bar{\nu}^c_{2 L}N^c_{2R}+
v_{s}\bar{N}_{2R}\nu_{2L}+\La_{s}\bar{N}_{2R}N^c_{2R})\crn &&+\fr
1 2 x(\la_{s}\bar{\nu}^c_{3 L}\nu_{3L}+v_{s}\bar{\nu}^c_{3
L}N^c_{3R}+v_{s}\bar{N}_{3R}\nu_{3L}+\La_{s}\bar{N}_{3R}N^c_{3R})\crn
&&+\fr 1 2 y(\la_{s}\bar{\nu}^c_{1 L}\nu_{2L}+ v_{s}\bar{\nu}^c_{1
L}N^c_{2R}+v_{s}\bar{N}_{1R}\nu_{2L}+\La_{s}\bar{N}_{1R}N^c_{2R})\crn
&&+\fr 1 2 y(\la_{s}\bar{\nu}^c_{1 L}\nu_{3L}+v_{s}\bar{\nu}^c_{1
L}N^c_{3R}+v_{s}\bar{N}_{1R}\nu_{3L}+\La_{s}\bar{N}_{1R}N^c_{3R})\crn
&& +\fr 1 2 x' (\la'_s\bar{\nu}^c_{2 L}\nu_{2L}+v'_s\bar{\nu}^c_{2
L}N^c_{2R}+
v'_s\bar{N}_{2R}\nu_{2L}+\La'_s\bar{N}_{2R}N^c_{2R})\crn &&+\fr 1
2 y'(\la'_s\bar{\nu}^c_{1 L}\nu_{3L}+v'_s\bar{\nu}^c_{1
L}N^c_{3R}+v'_s\bar{N}_{1R}\nu_{3L}+\La'_s\bar{N}_{1R}N^c_{3R})\crn
&& + \fr 1 2 z v_\rho (-\bar{\nu}^c_{2L}
N^c_{3R}+\bar{N}_{2R}\nu_{3L}+\bar{\nu}^c_{3L}
N^c_{2R}-\bar{N}_{3R}\nu_{2L}) \crn &&+ h.c. \label{s3nm1t4}\eea
We can rewrite \bea -\mathcal{L}^{\mathrm{mass}}_\nu=\fr 1 2
\bar{\chi}^c_L M_\nu \chi_L+ h.c.,\hs  \chi_L\equiv
\left(%
\begin{array}{c}
  \nu_L \\
  N^c_R \\
\end{array}%
\right),\hs M_\nu\equiv\left(%
\begin{array}{cc}
  M_L & M^T_D \\
  M_D & M_R \\
\end{array}%
\right),\label{nm}\eea where
$\nu_L=(\nu_{1L},\nu_{2L},\nu_{3L})^T$,
$N_R=(N_{1R},N_{2R},N_{3R})^T$ and
\bea M_{L,R,D}=\left(%
\begin{array}{ccc}
0 & b_{1L,R,D} & b_{2L,R,D} \\
  b_{1L,R,D} & c_{1L,R,D} & d_{L,R,D} \\
  b_{2L,R,D} & -d_{L,R,D} & c_{2L,R,D} \\
\end{array}%
\right),\eea where \bea
 b_{1L} & =
&\frac{\la_s y}{2},\hs b_{1D}= \frac{v_{s}y}{2},\hs b_{1R} =
\frac{\La_{s}y}{2},\crn b_{2L} & =
&\frac{(\la_{s}y+\la'_{s}y')}{2},\hs
b_{2D}=\frac{(v_{s}y+v'_{s}y')}{2},\hs
 b_{2R} = \frac{(\La_{s}y+\La'_{s}y')}{2},\crn
 c_{1L} & = &
(\la_{s}x+\la'_{s}x'),\hs c_{1D}=(v_{s}x+v'_{s}x'),\hs
c_{1R}=(\La_{s }x+\La'_{s}x'),\crn c_{2L} & = & \la_{s}x,\hs
c_{2D}=v_{s }x,\hs c_{2R}=\La_{s }x,\crn d_{L}&=&d_R=0,\hs d_D = z
v_\rho\equiv d. \label{bL,cL}\nn \eea

The remarks are in order\begin{itemize}
\item The VEVs with even lepton-parity are: $\la_s$, $\la'_s$, $\La_s$ and $\La'_s$;
\item The VEVs with odd lepton-parity are: $v_s$, $v'_s$ and
$v_\rho$.
\end{itemize} If the lepton parity is conserved we have $M_D=0$ since
$v_s=v'_s=v_\rho=0$. There is no mixing between the left-handed
neutrinos and the neutral fermions. The observed neutrinos are
just $\nu_L$ with masses given by $M_L$ consisting of $\la_s$ and
$\la'_s$ VEVs which are naturally small as given in eV order in
similarity to the case of the standard model with scalar triplets
\cite{malv} (called type II seesaw mechanism \cite{seesaw2}).
However, this situation as we see below cannot fit the data
provided that the contribution of $s'$ is as a small perturbation.

In general with the combined cases and lepton parity breaking,
three observed neutrinos gain masses via a cooperation of type I
and type II seesaw mechanisms derived from (\ref{nm}) as \bea
M_{\mathrm{eff}}=M_L-M_D^TM_R^{-1}M_D=\left(%
\begin{array}{ccc}
  A & B_1 & B_2 \\
  B_1 & C_1 & \mathcal{D} \\
  B_2 & \mathcal{D} & C_2 \\
\end{array}%
\right), \label{Mef}\eea where \bea A&=&-\left(b_{1R}
b_{2D}-b_{1D}b_{2R}\right)^2/\left(b^2_{2R}c_{1R}+b^2_{1R}c_{2R}\right),\crn
B_1&=&\left[b_{1L}c_{1R}b^2_{2R}+b_{1L}c_{2R}b_{1R}^2+b_{1R}b_{2R}b_{2D}c_{1D}
-b_{1D}c_{1D}b^2_{2R}-b_{1D}b_{2D}c_{1R}b_{2R}\right.\crn
&&\left.-b_{1R}c_{2R}b^2_{1D}+db_{1R}(b_{1R}b_{2D}-b_{2R}b_{1D})\right]/
\left(b^2_{2R}c_{1R}+b^2_{1R}c_{2R}\right),\crn
B_2&=&\left[b_{2L}c_{1R}b^2_{2R}+b_{2L}c_{2R}b_{1R}^2+b_{1R}b_{2R}b_{1D}c_{2D}
-b_{2D}c_{2D}b^2_{1R}-b_{1D}b_{2D}c_{2R}b_{1R}\right.\crn
&&\left.-b_{2R}c_{1R}b^2_{2D}+db_{2R}(b_{1R}b_{2D}-b_{2R}b_{1D})\right]/
\left(b^2_{2R}c_{1R}+b^2_{1R}c_{2R}\right),\crn
C_1&=&\left[b^2_{1D}c_{1R}c_{2R}+b^2_{2R}c_{1L}c_{1R}+b^2_{1R}c_{1L}c_{2R}
-2b_{1D}c_{1D}b_{1R}c_{2R}-c^2_{1D}b^2_{2R}\right.\crn
&&\left.-2db_{2R}(b_{1R}c_{1D}-b_{1D}c_{1R})-d^2b^2_{1R}\right]/\left(b^2_{2R}c_{1R}+b^2_{1R}c_{2R}\right),\crn
C_2&=&\left[b^2_{2D}c_{1R}c_{2R}+b^2_{1R}c_{2L}c_{2R}+b^2_{2R}c_{1R}c_{2L}
-2b_{2D}c_{2D}b_{2R}c_{1R}-c^2_{2D}b^2_{1R}\right.\crn
&&\left.-2db_{1R}(b_{2D}c_{2R}-b_{2R}c_{2D})-d^2b^2_{2R}\right]/\left(b^2_{2R}c_{1R}+b^2_{1R}c_{2R}\right),\crn
\mathcal{D}&=&\left[(b_{1D}c_{1R}-b_{1R}c_{1D})(b_{2D}c_{2R}-b_{2R}c_{2D})+d(b_{2D}b_{2R}c_{1R}
-b_{1D}b_{1R}c_{2R}\right.\crn
&&\left.+b_{1R}^2c_{2D}-b^2_{2R}c_{1D})-d^2b_{1R}b_{2R})\right]/\left(b^2_{2R}c_{1R}+b^2_{1R}c_{2R}\right).
\label{BCD}\eea The comments are in order: \begin{itemize} \item
If the subgroup $Z_2$ is unbroken, we have $A=\mathcal{D}=0$,
$B_1=B_2$ and $C_1=C_2$; \item If the $Z_2$ is broken by only the
case 1, we have $A=0$, $B_1=B_2$, $C_1=C_2$, but $\mathcal{D}\neq
0$; \item If the $Z_2$ is broken by both the cases, but the case 2
is regarded as a small perturbation, we have $A\approx 0$,
$B_1\approx B_2$, $C_1 \approx C_2$, and $\mathcal{D}\neq
0$.\end{itemize} In addition, it is able to introduce one more
antisextet in $\underline{1}$ which does not break the subgroup
$Z_2$ and implies $A\neq 0$ for contribution to our results above,
but this does not change our conclusions below and should be skip
without loss of generality.

We next divide our considerations into two cases to fit the data:
the first case is only case 1 above, and the second case is a
combination of the both.

\subsection{Experimental constraints under case 1 only}

In the case 1, $\la'_s=v'_s=\La'_s=0$, we have $A=0,\
B_1=B_2\equiv B,\ C_1 = C_2\equiv C,\  \mathcal{D}\equiv D\neq 0,$
and \bea M_{\mathrm{eff}}=\left(\begin{array}{ccc}
  0 & B & B \\
  B & C & D \\
  B & D & C \\
\end{array}%
\right),\label{dong2}\eea where \bea
B=\left(\la_s-\fr{v^2_s}{\La_s}\right)\fr{y}{2},\hs C=\left(\la_s
 - \fr{v^2_s}{\La_s}\right)x-\fr{v^2_{\rho}}{\La_s}\fr{z^2}{2 x},
 \hs D=-\fr{v^2_{\rho}}{\La_s}\fr{z^2}{2x}.\label{dong1}\eea
This mass matrix takes the form similar to that of the unbroken
$Z_2$ (with $v_{\rho}=0$). However, the breaking of $Z_2$
($v_\rho\neq 0$, thus $D\neq 0$) in this case is necessary to fit
the data (see below). Indeed, we can diagonalize $U^T
M_{\mathrm{eff}}U=\mathrm{diag}(m_1,m_2,m_3)$ where \bea m_1
&=&\fr 1 2 \left(C + D - \sqrt{8 B^2 + (C + D)^2}\right),\crn
m_2&=&\fr 1 2 \left(C + D + \sqrt{8 B^2 + (C + D)^2}\right),\label{dd7}\\
m_3&=&C-D,\nn\eea and the corresponding eigenstates put in the
lepton mixing matrix: \be U= \left(%
\begin{array}{ccc}
  -m_2/\sqrt{m^2_2+2B^2} & -m_1/\sqrt{m^2_1+2B^2} & 0 \\
  B/\sqrt{m^2_2+2B^2} & B/\sqrt{m^2_1+2B^2} & -\fr{1}{\sqrt{2}} \\
  B/\sqrt{m^2_2+2B^2} & B/\sqrt{m^2_1+2B^2} & \fr{1}{\sqrt{2}} \\
\end{array}%
\right).\label{dd8}\ee Note that $m_1m_2=-2B^2$. This matrix can
be parameterized in three Euler's angles, which implies: \be
\theta_{13}=0,\hs \theta_{23}=\pi/4,\hs \tan
\theta_{12}=\sqrt{-\fr{m_1}{m_2}}.\label{ntrbi}\ee

This case coincides with the data since $\theta_{13}<\pi/13$ and
$\theta_{23}\simeq \pi/4$ \cite{pdg} and close to the proposal of
the tribimaximal neutrino-mixing form \cite{hps}. For the
remaining constraints, taking the central values from the data
\cite{pdg} as displayed in (\ref{olddata}), $t^2_{12}\simeq 0.435$
(i.e. $s^2_{12}=0.304$), $\Delta m^2_{21}=7.65\times 10^{-5}\
\mathrm{eV}^2$ and $|\Delta m^2_{31}|=2.4\times 10^{-3}\
\mathrm{eV}^2$, we have a solution $m_1=-0.42\times 10^{-2}\
\mathrm{eV}$, $m_2=0.97\times 10^{-2}\ \mathrm{eV}$ and
$m_3=4.9\times 10^{-2}\ \mathrm{eV}$ (normal ordering). It follows
$B\simeq 0.451\times 10^{-2}$~eV, $C\simeq 2.725\times 10^{-2}$~eV
and $D\simeq -2.175\times 10^{-2}$~eV (as expected). Now, it is
natural to choose $\la_s$, $v^2_s/\La_s$ and $v^2_{\rho}/\La_s$ in
eV order. From (\ref{dong1}), we can find the three parameters
$x$, $y$ and $z$, with the provided $B$, $C$ and $D$,
respectively. It is noteworthy that in this case the contribution
of the $Z_2$ (or $S_3\rightarrow Z_3$) breaking parameter ($\sim
v^2_\rho/\La_s$) transforming under $\underline{1}'$ to the
neutrino masses is comparable to that of the $S_3\rightarrow Z_2$
one ($\sim \la_s,\ v^2_s/\La_s$). This makes the model viable and
in some sense quite similar to those of the quark and
charged-lepton sectors.

The recent considerations have implied $\theta_{13}\neq 0$
\cite{mal,flivale}, but small as given in (\ref{newdata}). If it
is correct, this case will fail. But, a combination of the case 1
with the case 2 above improves this.

\subsection{Experimental constraints under combination of case 1 and case 2}

In a scenario where both the case 1 and case 2 are taken place,
the neutrino mass matrix (\ref{Mef})
can be rewritten in the form: \be M_{\mathrm{eff}}=\left(%
\begin{array}{ccc}
  0 & B & B \\
  B & C & D \\
  B & D & C \\
\end{array}%
\right)+ \left(%
\begin{array}{ccc}
  \epsilon & p_1 & p_2 \\
  p_1 & q_1 & r \\
  p_2 & r & q_2 \\
\end{array}%
\right),\label{ddd}\ee where $B$, $C$ and $D$ are given by
(\ref{dong1}) accommodated in the first term or matrix due to the
contribution of the scalar antisextet $s$ and triplet $\rho$ only
as in the case 1 (\ref{dong2}). The last matrix is a deviation
from the contribution of the case 1 due to the scalar antisextet
$s'$, namely $\epsilon\equiv A$, $p_{1,2}=B_{1,2}- B$,
$q_{1,2}=C_{1,2} - C$ and $r=\mathcal{D}-D$, with the $A$,
$B_{1,2}$, $C_{1,2}$ and $\mathcal{D}$ being defined in
(\ref{BCD}). Indeed, if the case 2 or the $s'$ contribution is
forbidden, the deviations $\epsilon,\ p,\ q,\ r$ will vanish (to
be concrete, see below), therefore the mass matrix
$M_{\mathrm{eff}}$ (\ref{ddd}) reduces to its first term
coinciding with the case~1. The first term as shown can
approximately fit the new data \cite{flivale} with a ``small''
deviation for $\theta_{13}$. The second term proportional to
$\epsilon,\ p,\ q,\ r$ due to contribution of the antisextet $s'$
will take the role for such a deviation of $\theta_{13}$. So, in
this case we consider the $s'$ contribution as a small
perturbation and terminating the theory at the first order.

Provided that $\langle s' \rangle \ll \langle s \rangle $ or
$\la'_s/\la_s\sim v'_s/v_s\sim\La'_s/\La_s\ll 1$, one can evaluate
\be \epsilon \simeq -\fr{y'^2 v^2_s }{8x
\La_s}\left(\fr{v'_s}{v_s}-\fr{\La'_s}{\La_s}\right)^2\ll 1,\ee
which lies at the second order of the perturbation, thus ignored.
The remaining parameters $p_{1,2},\ q_{1,2},\ r$ are easily
obtained as follows \bea
p_1&\simeq&\frac{1}{16x\La_s}\left\{\left[\La_s\la_s(x'y+2xy')
-\fr{v^2_sx'(3x+2y)}{3}+v_sv_\rho
y'z\right]\frac{\La'_s}{\La_s}+\fr{v_s}{3}(v_sx'y+3v_\rho
y'z)\frac{v'_s}{v_s}\right\}, \crn
p_2&\simeq&\frac{1}{16x\La_s}\left\{\left[\La_s\la_s(x'y+2xy')
-\fr{v^2_sx'(x-3y)}{3}+v_sv_\rho
y'z\right]\frac{\La'_s}{\La_s}+\La_s\la_sx(x'+y')\frac{\la'_s}{\la_s}\right.\crn&&
+\left.\fr{v_s}{3}y'(3v_\rho
z-v_sx)\frac{v'_s}{v_s}\right\},\crn
q_1&\simeq&\frac{1}{8x\La_s}\left\{\left[\frac{v^2_sx}{y}(x'y-2xy')
+\frac{\La_s\la_s x}{y}(x'y+2xy')+2v_sv_\rho
x'z\right]\frac{\La'_s}{\La_s}+\La_s\la_sx'(x+x')\frac{\la'_s}{\la_s}\right.\crn&&-\left.
2v_sx'\left[v_s(x+z)+v_\rho z\right]\frac{v'_s}{v_s}\right\},\crn
q_2&\simeq&\frac{1}{8\La_s}\left\{\left[\La_s\la_s\frac{(x'y+2xy')}{y}+\frac{v_sy'}{y}(4v_\rho
z-v_sx)-\frac{2v^2_\rho
y'z^2}{xy}\right]\frac{\La'_s}{\La_s}+\frac{v_sy'}{y}(v_sx-4zv_\rho)\frac{v'_s}{v_s}\right\},\crn
r&\simeq&\frac{1}{8x\La_s}\frac{v_\rho
z}{y}\left\{\left[\la_s(x'y-xy')-v_\rho
y'z\right]\frac{\La'_s}{\La_s}+v_s(xy'-x'y)\frac{v'_s}{v_s}\right\},\eea
which all start from the first order of the perturbation. {\it It
is noteworthy that the presence of $\rho$ is important since by
contrast the $r$ will start from the second order and ignored as
$\epsilon$. The resulting mass matrix cannot fit the new data.}

The explicit form of the mass matrix (\ref{ddd}) is thus given by
\be M_{\mathrm{eff}}=\left(%
\begin{array}{ccc}
  0 & B & B \\
  B & C & D \\
  B & D & C \\
\end{array}%
\right)+ \fr{\la'_s}{\la_s}M_\la
+\fr{v'_s}{v_s}M_v+\fr{\La'_s}{\La_s} M_\La, \ee where the last
three terms are the perturbative contributions at the first order
with
\bea M_\la &\equiv& \fr{1}{8x}\left(%
\begin{array}{ccc}
  0 & 0 & \fr{\la_sx(x'+y')}{2} \\
  0 & \la_sx'(x+x') & 0 \\
  \fr{\la_sx(x'+y')}{2} & 0 & \la_sx'(x+x') \\
\end{array}%
\right),\crn M_v &\equiv& \fr{v_s}{8x\La_s}\left(%
\begin{array}{ccc}
  0 & \fr{1}{6}(v_sx'y+3v_\rho
y'z) & \fr{1}{6}y'(3v_\rho
z-v_sx) \\
  \fr{1}{6}(v_sx'y+3v_\rho
y'z) & 2x'\left[v_s(x+z)+v_\rho z\right] & \frac{v_\rho
z}{y}(xy'-x'y) \\
  \fr{1}{6}y'(3v_\rho
z-v_sx) & \frac{v_\rho
z}{y}(xy'-x'y) & \frac{y'}{y}(v_sx-4zv_\rho) \\
\end{array}%
\right),\crn
M_\La &\equiv & \fr{1}{8x\La_s}\left(%
\begin{array}{c}
 0  \\
  \fr{1}{2}\left[\La_s\la_s(x'y+2xy')-\fr{v^2_sx'(3x+2y)}{3}+v_sv_\rho
y'z\right]  \\
  \fr{1}{2}\left[\La_s\la_s(x'y+2xy')-\fr{v^2_sx'(x-3y)}{3}+v_sv_\rho
y'z\right]  \\
\end{array}%
\right.\crn
&&\hspace{1.3cm}\left.%
\begin{array}{c}
  \fr{1}{2}\left[\La_s\la_s(x'y+2xy')-\fr{v^2_sx'(3x+2y)}{3}+v_sv_\rho
y'z\right]  \\
  \frac{v^2_sx}{y}(x'y-2xy')+\frac{\La_s\la_s}{y}(x'y+2xy')+2v_sv_\rho
x'z  \\
   \frac{v_\rho z}{y}\left[\la_s(x'y-xy')-v_\rho
y'z\right]  \\
\end{array}%
\right.\crn
&&\hspace{1.3cm}\left.%
\begin{array}{c}
 \fr{1}{2}\left[\La_s\la_s(x'y+2xy')-\fr{v^2_sx'(x-3y)}{3}+v_sv_\rho
y'z\right] \\
   \frac{v_\rho z}{y}\left[\la_s(x'y-xy')-v_\rho
y'z\right] \\
  \La_s\la_s\frac{(x'y+2xy')}{y}+\frac{v_sy'}{y}(4v_\rho
z-v_sx)-\frac{2v^2_\rho
y'z^2}{xy} \\
\end{array}%
\right),\eea which all are in the same order with the dominant
contributions $B,C,D$ proportional to the mass scale of observed
neutrinos $\la_s,\ v^2_s/\La_s,\ v^2_\rho/\La_s,\ v_s
v_\rho/\La_s$.

The physical neutrino masses at the first order are obtained as
\bea m'_{1,2}&=&
m_{1,2}+\frac{B^2}{(m_{2,1})^2+2B^2}\frac{1}{8xy\La_s}\left(K^1_{1,2}\fr{\La'_s}{\La_s}
+K^2_{1,2}\fr{\la'_s}{\la_s}+K^3_{1,2}\fr{v'_s}{v_s}\right),\crn
m'_3&=&m_3+\frac{1}{8xy\La_s}\left(K^1_3\fr{\La'_s}{\La_s}+K^2_3\fr{\la'_s}{\la_s}+K^3_3\fr{v'_s}{v_s}\right),
\eea where \bea K^1_1&=&-3v^2_\rho
y'z^2+v_s^2x(x'y-3xy')+2v_\rho\la_s z(x'y-x y')+ 2v_sv_\rho
z(x'y+2xy')\crn&& +2\la_s\La_s x(x'y+2xy')-\fr{2m_2}{B}[\La_s\la_s
y(x'y+2xy')-\fr{v^2_sx'y(4x-y)}{6}+v_sv_\rho yy'z],\crn
K^2_1&=&\La_s\la_sy\left[x'(x+x')-\fr{m_2}{B}x(x+x')\right],\crn
K^3_1&=&v_s[-2v_sx'y(x+z)+v_\rho
z(-3yx'+xy')]-\frac{m_2}{B}\frac{v_sy}{3}[v_s(x'y-xy')+6v_\rho
y'z],\crn
K^1_3&=&-\fr{v^2_\rho y'z^2}{2}+\fr{v^2_s x(x'y-3x
y')}{2}+v_\rho\la_s z(-x'y+xy')+v_sv_\rho z(x'y+2xy')\crn
&&+\la_s\La_s x(x'y+2xy'),\crn
K^2_3&=&\La_s\la_s x'y(x+x'),\hs K^3_3=-v_s[2v_sx'y(x+z)+v_\rho
z(x'y+xy')], \eea with $K^a_2$ $(a=1,2,3)$ similarly given as
$K^a_1$ but $m_2$ is replaced by $m_1$. The $m_{1,2,3}$ are the
mass values as of the case 1 given by (\ref{dd7}). For the
corresponding perturbed eigenstates, we put \be U\longrightarrow
U'= U+\Delta U,\ee where $U$ is defined by (\ref{dd8}) as of the
case 1 and
\be \Delta U= \left(%
\begin{array}{ccc}
  -\fr{m_1}{\sqrt{m^2_1+2B^2}}F_1 & \fr{m_2}{\sqrt{m^2_2+2B^2}}F_1 &
  \fr{m_1}{\sqrt{m^2_1+2B^2}}F_3-\fr{m_2}{m^2_1+2B^2}F_2 \\
  \fr{B}{\sqrt{m^2_1+2B^2}}F_1+\fr{F_2}{2} & -\fr{B}{\sqrt{m^2_2+2B^2}}F_1-\fr{F_3}{2}
  & \fr{B}{\sqrt{m^2_2+2B^2}}F_2-\fr{B}{\sqrt{m^2_1+2B^2}}F_3 \\
  \fr{B}{\sqrt{m^2_1+2B^2}}F_1-\fr{F_2}{2} & -\fr{B}{\sqrt{m^2_2+2B^2}}F_1+\fr{F_3}{2}
  & \fr{B}{\sqrt{m^2_2+2B^2}}F_2-\fr{B}{\sqrt{m^2_1+2B^2}}F_3 \\
\end{array}%
\right),\label{PMNS2}\ee with \bea
F1&=&\frac{(m_1+m_2)(p_1+p_2)-B(q_1+q_2+2r)}{\sqrt{2}(m_1-m_2)^2},\crn
F2&=&\frac{m_2(p_2-p_1)+B(q_1-q_2)}{(m_1-m_3)\sqrt{m^2_2+2B^2}},\crn
F3&=&\frac{m_1(p_2-p_1)+B(q_1-q_2)}{(m_2-m_3)\sqrt{m^2_1+2B^2}}.
\eea The lepton mixing matrix in this case $U'$ can still be
parameterized in three new Euler's angles $\theta'_{ij}$, which
are also a perturbation from the $\theta_{ij}$ in the case 1,
defined by \bea s'^2_{13}&=&\left|\sqrt{\fr{1}{1+\al^2}}(\al
F_3-F_2)\right|^2,\crn
t'^2_{12}&=&\left|\al(1-\fr{1+\al^2}{\al}F_1)\right|^2,\crn
t'^2_{23}&=&\left|\fr{1}{\sqrt{2}}-2\sqrt{\fr{1}{1+\al^2}}(\al
F_2-F_3)\right|^2,\nn \eea where $\al=\sqrt{-\fr{m_1}{m_2}}$ is
just the $t_{12}$ of the case 1.

It is easily shown that our model is consistent since the five
experimental constraints on the mixing angles and squared mass
differences of neutrinos can be respectively fitted with the five
Yukawa coupling parameters $x,\ y,\ x',\ y',\ z$ of the $s$, $s'$
antisextets and $\rho$ triplet scalars, provided that the VEVs are
previously given. Indeed, let us first assume
$\la_s=v^2_s/\La_s=v_sv_{\rho}/\La_s = v^2_{\rho}/\La_s= 1\
\mathrm{eV}$ and $\la'_s/\la_s = v'_s/v_s = \La'_s/\La_s = 0.1$
which are safely small. Taking the new data (\ref{newdata}):
$s'^2_{13}= 0.013$, $s'^2_{12}= 0.312$, $s'^2_{23}= 0.52$ as well
as $\Delta m'^2_{21}=7.59\times 10^{-5}\ \mathrm{eV}^2$ and
$|\Delta m'^2_{31}|=2.5\times 10^{-3}\ \mathrm{eV}^2$, we obtain
$x\simeq 0.049$, $y\simeq 0.00895$, $x'\simeq 0.00149$, $y'\simeq
9.02\times 10^{-4}$ and $z\simeq 0.0461 $. The neutrino masses are
explicitly given as $m'_1=-0.41\times 10^{-2}\ \mathrm{eV}$,
$m'_2=0.97\times 10^{-2}\ \mathrm{eV}$ and $m'_3=4.9\times
10^{-2}\ \mathrm{eV}$ which are in a normal ordering.

\subsection{Remark on breaking, VEVs and rho parameter}

Both the fitting cases mentioned above require $\mathcal{D}\neq
0$. Hence, to have a consistent neutrino spectrum we conclude that
the lepton parity must be broken because by contrast $\mathcal{D}$
vanishes. Also, both the directions $S_3\rightarrow Z_2$ and
$S_3\rightarrow Z_3$ must be taken place.

We remark that for both the fitting cases the seesaw scale $\La_s$
is not needed to be so large that can naturally be taken at TeV
scale as the VEV $\om$ of $\chi$. This is because $v_s$ and
$v_\rho$ carry lepton number, simultaneously breaking the lepton
parity which are naturally constrained to be much smaller than the
electroweak scale \cite{manf,dlshA4,dlsvS4}. This is also behind a
theoretical fact that $\om$, $\La_s$ are scales for the gauge
symmetry breaking in a stage from $\mathrm{SU}(3)_L\otimes
\mathrm{U}(1)_X\rightarrow \mathrm{SU}(2)_L\otimes
\mathrm{U}(1)_Y$. They will provide masses for the new gauge
bosons such as $Z'$, $X$ and $Y$. Also, the exotic quarks gain
masses from $\om$ while the neutral fermions are from $\La_s$. The
second stage of the gauge symmetry breaking is from
$\mathrm{SU}(2)_L\otimes \mathrm{U}(1)_Y\rightarrow
\mathrm{U}(1)_Q$ achieved by the electroweak scale VEVs such as
$u,u',v,v'$ responsible for ordinary particle masses. In
combination with those of type II seesaw as determined, the
hierarchies in VEVs are summarized as \be \mathrm{eV} \sim \la_s <
v_s,\ v_\rho < v,\ v',\ u,\ u'< \La_s,\ \om\sim
\mathrm{TeV}.\label{vevss}\ee Here the VEVs of $s'$ as the role of
perturbation, $\la'_s/\la_s\sim v'_s/v_s\sim \La'_s/\La_s \ll 1$,
are not mentioned.

Our model contains a lot of $\mathrm{SU}(2)_L$ scalar doublets and
triplets that may modify the precision electroweak data. The most
serious one can result from the tree-level contributions to the
$\rho$ parameter. To see this let us approximate $W$ mass and
$\rho$: \bea m^2_W = \frac{g^2}{2}v^2_w,\hs \rho=\fr{m^2_W}{c^2_W
m^2_Z}\simeq 1-\fr{2\lambda^2_s}{v^2_w}, \eea where $v^2_w\simeq
(u^2+u'^2+v^2+v'^2)= (174\, \textrm{GeV})^2$ is naturally given
due to (\ref{vevss}) and $ \langle\chi^0_1\rangle\ll u,\ u',\ v,\
v'$ by the same reason as $v_\rho,\ v_s$. Because $\lambda_{s}$ is
in eV scale responsible for the observed neutrino masses, the
$\rho$ is absolutely close to one and in agreement with the data
\cite{pdg}.

\section{\label{331RH} $S_3$ symmetry in the 3-3-1 model with right-handed neutrinos ($\nu_R$)}

The fermion content of this model can be given as \bea \psi_{1L}
&=& \left(
    \nu_{1L},\ l_{1L},\ \nu^c_{1R}\right)^T\sim
    [3,-1/3,1/3,\underline{1}],\hs
    l_{1R}\sim[1,-1,1,\underline{1}],
    \crn \psi_{ \alpha L }&=&
\left(\nu_{\alpha L},\ l_{\alpha L},\ \nu^c_{\alpha R} \right)^T
\sim [3,-1/3,1/3,\underline{2}],\hs l_{\alpha
R}\sim[1,-1,1,\underline{2}], \crn
 Q_{1L}&=& \left(u_{1L},\ d_{1L},\
 U_{L}\right)^T\sim[3,1/3,-2/3,\underline{1}],\crn
u_{1R} &\sim &[1,2/3,0,\underline{1}],\hs
d_{1R}\sim[1,-1/3,0,\underline{1}],\hs U_R\sim
[1,2/3,-2,\underline{1}],\label{conts3}\\
 Q_{\al L}&=&\left(d_{\alpha L},\ -u_{\alpha L},\ D_{\alpha
 L}\right)^T\sim[3^*,0,2/3,\underline{2}],\crn
u_{\al R}&\sim& [1,2/3,0,\underline{2}],\hs d_{\al
R}\sim[1,-1/3,0,\underline{2}],\hs D_{\al R}
\sim[1,-1/3,2,\underline{2}].\nn\eea Here the difference from the
previous model is that $L(\nu_R)=1$. Hence the exotic quarks and
new non-Hermitian gauge bosons are bilepton:
$L(D)=-L(U)=L(X^0)=L(Y^-)=2$. And, the leptonic operator is given
by $L=\fr{4}{\sqrt{3}}T_8+\mathcal{L}$ with $\mathcal{L}$ listed
in the square brackets.

The scalar sector in this model is similar to, but simpler than
the previous model since the $\rho$ triplet is not necessary to be
introduced. This is due to the fact that the interaction
$(\bar{\psi}^c_{ 2L} \psi_{3 L}-\bar{\psi}^c_{3L} \psi_{2
L})\phi'$ is allowed (since $\mathcal{L}$ is conserved), thus the
$\phi'$ triplet can play the role instead of $\rho$ as in the case
1 above. Notice that the $\mathcal{L}$ charges of all the scalars
in this model are changed while the other charges are remained:
\bea \phi &\sim& [3,2/3,-2/3, \underline{1}],\hs \phi' \sim
[3,2/3,-2/3, \underline{1}'] \crn \eta &\sim&
[3,-1/3,-2/3,\underline{1}],\hs
\eta'\sim[3,-1/3,-2/3,\underline{1}'],\hs \chi \sim
[3,-1/3,4/3,\underline{1}],\crn
s&\sim&[6^*,2/3,-2/3,\underline{2}],\hs s'\sim
[6^*,2/3,-2/3,\underline{2}].\eea

The charged-lepton and quark masses are similar to the pervious
model. However, the lepton parity in this case does not work.
Exactly, it cannot suppress the mixing between ordinary quarks and
exotic quarks since the odd fields as mentioned in the previous
model are now even $-L(\eta^0_3)=-L(\eta'^0_3)=L(\chi^0_1)=2$
which can develop VEVs. The lepton numbers of particles in this
model are listed in App. \ref{apb}. This mixing can only be
prevented if we suppose that the $L$ charge is not spontaneously
broken. However, this discards the seesaw mechanism since
$\la_{s}=\la'_s=\La_s=\La'_s=0$. The neutrinos have only Dirac
masses via the VEVs $v'$ of $\phi'$, $v_s$ and $v'_s$ which are
not natural in the same simple extension of the standard model
with Dirac neutrino masses \cite{dls}.

Let us recall that it is different from the previous model since
in that case we have no mixing between the two kinds of quarks due
to the lepton parity while $L$ is still broken responsible for the
neutrino masses via the type I seesaw mechanism where
$v_\rho=v_s=v'_s=0$. However, as mentioned it is not realistic
under the data when the $s'$ contribution is regarded as a small
perturbation. In contradiction, if the $s'$ contribution becomes
comparable, the situation will change (which has not been
considered in the present work).

All the issues above can be resolved by imposing a spontaneous
symmetry breaking of $L$. This breaking can be explicitly derived
via a $L$-violating scalar potential. It also proves that the VEVs
$\la_s,\ v_s^2/\La_s,\ v_s v'/\La_s,\ v'^2/\La_s$ responsible for
the observed neutrino masses are naturally small
\cite{dlsvS4,dlshA4}. The results on the neutrino masses and
mixings are given in similarity to the previous model with the
replacement of $v_\rho$ by $v'$. The mixing between the exotic
quarks and ordinary quarks at the tree level happens but small for
the same reason. The flavor changing neutral current starts from
the tree level but strongly suppressed \cite{dlsvS4,dlshA4}.
However, the difference from the previous model is that since the
$v'$ and $v_s$ belong to the electroweak scale the seesaw scale
$\La_s$ is not needed to be in TeV order. In principle it can flip
up to a very high scale such as the GUT one.

\section{\label{conclus}Conclusions}

We have argued that the 3-3-1 models may accommodate the seesaw
mechanisms naturally. In fact, the right-handed neutrinos or
neutral fermions can exist as basic objects needed to complete
multiplets extended from those of the standard model. We have
shown that the TeV seesaw mechanism can be naturally obtained in
the 3-3-1 model with neutral fermions. Whereas, in the 3-3-1 model
with right-handed neutrinos the mechanism can work up to a very
high scale such as the GUT's. In our framework, a combination of
type I and II seesaws is always in the cooperation.

We have also argued that due to anomaly cancelation the 3-3-1
models may naturally permit of flavor symmetries such as $S_4$ and
$S_3$ which has been taken into account since they possess
$\underline{2}$ representations responsible for the 3-3-1 quark
sector and $\mu-\tau$ symmetry as known. In addition, the 3-3-1
models can work only with three families as the flavor symmetries
do. In the standard model, the families are in replication, thus
naturally to put all in $\underline{3}$ which is
$\underline{1}\oplus \underline{2}$ under $S_3$. By this
indication, we have put the first family in $\underline{1}$ and
the last two in $\underline{2}$ to realize successful mass spectra
and mixings for leptons and quarks.

We have introduced a new charge $\mathrm{U}(1)_{\mathcal{L}}$
responsible for lepton number and lepton parity. The two 3-3-1
models as given are already in difference due to
$\mathcal{L}$-charge embedding. In the 3-3-1 model with neutral
fermions, the $N_R$s have vanishing lepton number and the lepton
parity being in operation to realize the TeV seesaw mechanism in
similarity to the scenario previously proposed \cite{manf}. In the
3-3-1 model with right-handed neutrinos, the lepton parity cannot
work which realizes the popular seesaw mechanism. The scalar
sector for the two models is also in difference. We have briefly
discussed that if $\mathrm{U}(1)_{\mathcal{L}}$ is violated via
the scalar potential as given in Apps. \ref{apc}, the seesaw
contributions are generated to be naturally small, responsible for
the observed neutrino masses. By this reason the tree level exotic
and ordinary quark mixing and flavor changing neutral current are
also strongly suppressed. All those are in similarity to
\cite{manf,malv,dlsvS4,dlshA4}.

We have shown that the realistic neutrino mixing can be obtained
if the two directions for breaking $S_3\rightarrow Z_2$ and
$S_3\rightarrow Z_3$ (or $S_3\rightarrow Z_2\rightarrow
\{\mathrm{Indentity}\}$) simultaneously take place and equivalent
in size, i.e. the contributions due to $\rho$ (or $\phi'$) and $s$
are comparable. If the $s'$ which is also responsible for
$Z_2\rightarrow \{\mathrm{Indentity}\}$ breaking is not
introduced, the models can fit the old data with $\theta_{13}=0$.
Otherwise, if it is presented as a small perturbation in
contributing to the mass spectrum the new data under the light of
the new observations can be naturally recognized.

\section*{Acknowledgments}
This work was supported in part by the National Foundation for
Science and Technology Development of Vietnam (NAFOSTED).
\\[0.3cm]
\appendix


\section{\label{apa}$\emph{S}_3$ group and Clebsch-Gordan coefficients}

$S_3$ is the permutation group of three objects, having six
elements divided into three conjugacy classes \cite{kj}. We denote
\underline{1}, \underline{1}$'$, and \underline{2} as its three
irreducible representations, and $n$, $h$ as the order of class
and the order of elements within each class, respectively. The
character table is given by
\begin{center}
\begin{tabular}{|c|c|c|c|c|c|}
\hline Class & $n$ & $h$ & $\chi_{\underline{1}}$ &
$\chi_{\underline{1}'}$ & $\chi_{\underline{2}}$
\\
\hline
$C_1$ & 1 & 1 & 1 & 1 & 2 \\
$C_2$ & 2 & 3 & 1 & 1 & -1 \\
$C_3$ & 3 & 2 & 1 & -1 & 0 \\
\hline
\end{tabular}
\end{center}

We will work in a basis in which the representation
$\underline{2}$ is complex (See, for example, Ma in \cite{kj}).
The decomposition rules can be obtained as \bea
\underline{1}\otimes\underline{1}&=&\underline{1}(11),\hs
\underline{1}'\otimes \underline{1}'=\underline{1}(11),\hs
\underline{1}\otimes\underline{1}'=\underline{1}'(11),\crn
\underline{1}\otimes \underline{2}&=&\underline{2}(11,12),\hs
\underline{1}'\otimes \underline{2}=\underline{2}(11,-12),\crn
\underline{2} \otimes \underline{2} &=& \underline{1}(12+21)
\oplus \underline{1}'(12-21) \oplus \underline{2}(22,11),\nn \eea
where for the terms in parentheses the first and second factor
indicate to the multiplet components of the first and second
representations, respectively. The conjugation rules
 are given by \bea
\underline{2}^*(1^*,2^*)&=&\underline{2}(2^*,1^*),\hs
\underline{1}^*(1^*)=\underline{1}(1^*),\hs
\underline{1}'^*(1^*)=\underline{1}'(1^*).\nn\eea

\section{\label{apb}Lepton number}

For convenience, we also list the lepton number ($L$) of particles
for the two models as mentioned.

\subsection{The 3-3-1 model with neutral fermions}

\bc
\begin{tabular}{|c|c|}
  \hline
  Particle & $L$   \\ \hline
$u$, $d$, $N_R$, $W$, $Z$, $\phi^+_1$, $\phi^0_2$,  $\phi'^+_1$,
$\phi'^0_2$, $\eta^0_1$, $\eta^-_2$, $\eta'^0_1$, $\eta'^-_2$,
  $\chi^0_3$, $s^0_{33}$, $s'^0_{33}$ & 0  \\ \hline
   $\nu^*_L$, $l^*$, $U$, $D^*$, $X^{0*}$, $Y^+$, $\phi^+_3$, $\phi'^+_3$, $\eta^0_3$, $\eta'^0_3$, $\chi^{0*}_1$, $\chi^+_2$,
      $s^0_{13}$, $s^+_{23}$, $s'^0_{13}$, $s'^+_{23}$, $\rho^+_1$, $\rho^0_2$ & $-1$
   \\ \hline
   $s^{0}_{11}$, $s^{+}_{12}$, $s^{++}_{22}$, $s'^{0}_{11}$, $s'^{+}_{12}$, $s'^{++}_{22}$, $\rho^+_3$ & $-2$ \\ \hline
\end{tabular}\ec

\subsection{The 3-3-1 model with right-handed neutrinos}

\bc
\begin{tabular}{|c|c|}
  \hline
  Particle & $L$   \\ \hline
   $\nu_{L}$, $\nu_R$, $l$  & $1$ \\ \hline
$u$, $d$, $W$, $Z$, $\phi^+_1$, $\phi^0_2$,  $\phi'^+_1$,
$\phi'^0_2$, $\eta^0_1$, $\eta^-_2$, $\eta'^0_1$, $\eta'^-_2$,
  $\chi^0_3$, $s^0_{13}$, $s^+_{23}$, $s'^0_{13}$, $s'^+_{23}$ & 0
   \\ \hline
 $U$, $D^*$, $X^{0*}$, $Y^+$, $\phi^+_3$, $\phi'^+_3$, $\eta^0_3$, $\eta'^0_3$, $\chi^{0*}_1$,
 $\chi^+_2$, $s^{0*}_{33}$, $s'^{0*}_{33}$, $s^{0}_{11}$, $s^{+}_{12}$, $s^{++}_{22}$, $s'^{0}_{11}$,
 $s'^{+}_{12}$, $s'^{++}_{22}$ & $-2$ \\ \hline
\end{tabular}\ec

\section{\label{apc}Scalar potential}

To be complete, we write the scalar potentials of both the models
mentioned. It is also noted that $(\Tr{A})(\Tr{B})=\Tr{(A\Tr{B}})$
and
$V(\textit{X}\rightarrow\textit{X'},\textit{Y}\rightarrow\textit{Y'},\cdots)
\equiv V(X,Y,\cdots)\!\!\!\mid_{X=X',Y=Y',\cdots}$

\subsection{The 3-3-1 model with neutral fermions}

The general potential invariant under any group takes the form:
\be V_{\mathrm{total}}=V_{\mathrm{tri}}+V_{\mathrm{sext}}+
V_{\mathrm{tri-sext}},\label{vien5}\ee where $V_{\mathrm{tri}}$
comes from only contributions of $\mathrm{SU}(3)_L$ triplets given
as a sum of: \bea V(\chi)&=&\mu_{\chi}^2\chi^\+\chi
+\lambda^{\chi}({\chi}^\+\chi)^2,\label{Vchi}\\
V(\phi)&=&V(\chi\rightarrow\phi), \hs
V(\phi')=V(\chi\rightarrow\phi'),\label{vphiphip}\hs
V(\eta)=V(\chi\rightarrow\eta),\\
V(\eta')&=&V(\chi\rightarrow\eta'),\hs V(\rho)=V(\chi\rightarrow \rho),\label{vetaetaprho}\\
V(\phi,\chi)&=&\lambda_1^{\phi\chi}(\phi^\+\phi)(\chi^\+\chi)
+\lambda_2^{\phi\chi}(\phi^\+\chi)(\chi^\+\phi),\\
V(\phi,\phi')&=&V(\phi,\chi\rightarrow\phi')
+\lambda_3^{\phi\phi'}(\phi^\+\phi')(\phi^\+\phi')
+\lambda_4^{\phi\phi'}(\phi'^\+\phi)(\phi'^\+\phi),\\
V(\phi,\eta)&=&V(\phi,\chi\rightarrow\eta),\hs V(\phi,\eta')=V(\phi,\chi\rightarrow\eta'),\\
V(\phi,\rho)&=&V(\phi,\chi\rightarrow \rho),\hs
V(\phi',\chi)=V(\phi\rightarrow\phi',\chi),\\
V(\phi',\eta)&=&V(\phi\rightarrow\phi',\chi\rightarrow\eta),\hs
V(\phi',\eta')=V(\phi\rightarrow\phi',\chi\rightarrow\eta'),\\
V(\phi',\rho)&=&V(\phi\rightarrow\phi',\chi\rightarrow\rho),\hs
V(\chi,\eta)=V(\phi\rightarrow\chi,\chi\rightarrow\eta),\\
V(\chi,\eta')&=&V(\phi\rightarrow\chi,\chi\rightarrow\eta'),\hs V(\chi,\rho)
=V(\phi\rightarrow\chi,\chi\rightarrow\rho),\\
V(\eta,\eta')&=&V(\phi\rightarrow\eta,\chi\rightarrow\eta')
+\lambda_3^{\eta\eta'}(\eta^\+\eta')(\eta^\+\eta')+\lambda_4^{\eta\eta'}(\eta'^\+\eta)(\eta'^\+\eta),\\
V(\eta,\rho)&=&V(\phi\rightarrow\eta,\chi\rightarrow\rho),\hs
V(\eta',\rho)=V(\phi\rightarrow\eta',\chi\rightarrow\rho),\\
V_{\chi\phi\phi'\eta\eta'\rho}&=&\mu_1\chi\phi\eta+\mu'_1\chi\phi'\eta'
+\la^1_1(\phi^\+\phi')(\eta^\+
\eta')+\la^2_1(\phi^\+\phi')(\eta'^\+
\eta)+\la^3_1(\phi^\+\eta)(\eta'^\+ \phi')\crn
&&+\la^4_1(\phi^\+\eta')(\eta^\+\phi')
+\la^5_1(\phi^\+\rho)(\eta'^\+\chi)
+\la^6_1(\phi'^\+\rho)(\eta^\+\chi)+\la^7_1(\eta'^\+\rho)(\phi^\+\chi)\crn&&
+\la^8_1(\eta^\+\rho)(\phi'^\+\chi)+h.c.\label{vtrifourintract}
\eea The $V_{\mathrm{sext}}$ is summed from only antisextet
contributions: \bea V(s)&=&{\mu}^2_{s}\Tr(s^\+s)
+{\lambda}_1^{s}\Tr[(s^\+s)_{\underline{1}}(s^\+s)_{\underline{1}}]+
{\lambda}_2^{s}\Tr[(s^\+s)_{\underline{1'}}(s^\+s)_{\underline{1'}}]
+{\lambda}_3^{s}\Tr[(s^\+s)_{\underline{2}}(s^\+s)_{\underline{2}}]\crn
&+&{\lambda}_4^{s}\Tr(s^\+s)_{\underline{1}}\Tr(s^\+s)_{\underline{1}}
+{\lambda}_5^{s}\Tr(s^\+s)_{\underline{1'}}\Tr(s^\+s)_{\underline{1'}}
+{\lambda}_6^{s}\Tr(s^\+s)_{\underline{2}}\Tr(s^\+s)_{\underline{2}},\label{vs}\\
V(s')&=&V(s\rightarrow s'),\\
V(s,s')&=&{\mu}^2_{ss'}\Tr(s^\+s')+{\lambda}_1^{ss'}\Tr[(s^\+s)_{\underline{1}}(s'^\+s')_{\underline{1}}]+
{\lambda}_2^{ss'}\Tr[(s^\+s)_{\underline{1'}}(s'^\+s')_{\underline{1'}}]
+{\lambda}_3^{ss'}\Tr[(s^\+s)_{\underline{2}}(s'^\+s')_{\underline{2}}]\crn&&
+{\lambda}_4^{ss'}\Tr(s^\+s)_{\underline{1}}\Tr(s'^\+s')_{\underline{1}}
+{\lambda}_5^{ss'}\Tr(s^\+s)_{\underline{1'}}\Tr(s'^\+s')_{\underline{1'}}
+{\lambda}_6^{ss'}\Tr(s^\+s)_{\underline{2}}\Tr(s'^\+s')_{\underline{2}}\crn&&
+{\lambda}_7^{ss'}\Tr[(s^\+s')_{\underline{1}}(s'^\+s)_{\underline{1}}]+
{\lambda}_8^{ss'}\Tr[(s^\+s')_{\underline{1'}}(s'^\+s)_{\underline{1'}}]
+{\lambda}_9^{ss'}\Tr[(s^\+s')_{\underline{2}}(s'^\+s)_{\underline{2}}]\crn&&
+{\lambda}_{10}^{ss'}\Tr(s^\+s')_{\underline{1}}\Tr(s'^\+s)_{\underline{1}}
+{\lambda}_{11}^{ss'}\Tr(s^\+s')_{\underline{1'}}\Tr(s'^\+s)_{\underline{1'}}
+{\lambda}_{12}^{ss'}\Tr(s^\+s')_{\underline{2}}\Tr(s'^\+s)_{\underline{2}}\crn&&
+{\lambda}_{13}^{ss'}\Tr[(s^\+s')_{\underline{1}}(s^\+s')_{\underline{1}}]+
{\lambda}_{14}^{ss'}\Tr[(s^\+s')_{\underline{1'}}(s^\+s')_{\underline{1'}}]
+{\lambda}_{15}^{ss'}\Tr[(s^\+s')_{\underline{2}}(s^\+s')_{\underline{2}}]\crn&&
+{\lambda}_{16}^{ss'}\Tr(s^\+s')_{\underline{1}}\Tr(s^\+s')_{\underline{1}}
+{\lambda}_{17}^{ss'}\Tr(s^\+s')_{\underline{1'}}\Tr(s^\+s')_{\underline{1'}}
+{\lambda}_{18}^{ss'}\Tr(s^\+s')_{\underline{2}}\Tr(s^\+s')_{\underline{2}}\crn&&
+{\lambda}_{19}^{ss'}\Tr[(s^\+s)_{\underline{1}}(s^\+s')_{\underline{1}}]+
{\lambda}_{20}^{ss'}\Tr[(s^\+s)_{\underline{1'}}(s^\+s')_{\underline{1'}}]
+{\lambda}_{21}^{ss'}\Tr[(s^\+s)_{\underline{2}}(s^\+s')_{\underline{2}}]\crn&&
+{\lambda}_{22}^{ss'}\Tr(s^\+s)_{\underline{1}}\Tr(s^\+s')_{\underline{1}}
+{\lambda}_{23}^{ss'}\Tr(s^\+s)_{\underline{1'}}\Tr(s^\+s')_{\underline{1'}}
+{\lambda}_{24}^{ss'}\Tr(s^\+s)_{\underline{2}}\Tr(s^\+s')_{\underline{2}}\crn&&
+{\lambda}_{25}^{ss'}\Tr[(s'^\+s')_{\underline{1}}(s^\+s')_{\underline{1}}]+
{\lambda}_{26}^{ss'}\Tr[(s'^\+s')_{\underline{1'}}(s^\+s')_{\underline{1'}}]
+{\lambda}_{27}^{ss'}\Tr[(s'^\+s')_{\underline{2}}(s^\+s')_{\underline{2}}]\crn&&
+{\lambda}_{28}^{ss'}\Tr(s'^\+s')_{\underline{1}}\Tr(s^\+s')_{\underline{1}}
+{\lambda}_{29}^{ss'}\Tr(s'^\+s')_{\underline{1'}}\Tr(s^\+s')_{\underline{1'}}
+{\lambda}_{30}^{ss'}\Tr(s'^\+s')_{\underline{2}}\Tr(s^\+s')_{\underline{2}}\crn
&&+h.c. \eea The $V_{\mathrm{tri-sext}}$ is given as a sum of all
the terms connecting both the sectors: \bea
V(\phi,s)&=&\lambda_1^{\phi
s}(\phi^\+\phi)\Tr(s^\+s)_{\underline{1}} +\lambda_2^{\phi
s}[(\phi^\+s^\+)(s\phi)]_{\underline{1}},\\
V(\phi',s)&=&V(\phi\rightarrow\phi',s),\hs V(\chi,s)=V(\phi\rightarrow\chi,s),\hs V(\eta,s)=V(\phi\rightarrow\eta,s),\\
V(\eta',s)&=&V(\phi\rightarrow\eta',s),\hs V(\rho,s)=V(\phi\rightarrow\rho,s)+\{\la^{\rho s}_3\rho[(\rho s^\+)s^\+]_{\underline{1'}}+h.c\},\\
V(\phi,s')&=&V(\phi,s\rightarrow s'),\hs
V(\phi',s')=V(\phi\rightarrow\phi',s\rightarrow s'),\\
V(\chi,s')&=&V(\phi\rightarrow\chi,s\rightarrow s'),\hs
V(\eta,s')=V(\phi\rightarrow\eta,s\rightarrow s'),\\ V(\eta',s')&=&V(\phi\rightarrow\eta',s\rightarrow s'),\\
V(\rho,s')&=&V(\phi\rightarrow\rho,s\rightarrow s')+\{\la^{\rho s'}_3\rho[(\rho s'^\+)s'^\+]_{\underline{1'}}+h.c\},\\
V(\phi,s,s')&=&\lambda_1^{\phi
ss'}(\phi^\+\phi)\Tr(s^\+s')_{\underline{1}} +\lambda_2^{\phi
ss'}[(\phi^\+s^\+)(s'\phi)]_{\underline{1}}+h.c,\\
V(\phi',s,s')&=&V(\phi\rightarrow\phi',s,s'),\hs
V(\chi,s,s')=V(\phi\rightarrow\chi,s,s'),\\
V(\eta,s,s')&=&V(\phi\rightarrow\eta,s,s'),\hs
V(\eta',s,s')=V(\phi\rightarrow\eta',s,s'),\\ V(\rho,s,s')&=&V(\phi\rightarrow\rho,s,s')+\{\la^{\rho ss'}_3\rho[(\rho s^\+)s'^\+]_{\underline{1'}}+\la^{\rho ss'}_4\rho[(\rho s'^\+)s^\+]_{\underline{1'}}+h.c\},\\
V_{ss'\chi\phi\phi'\eta\eta'\rho}&=&
(\lambda_1\phi^\+\phi'+\lambda_2\eta^\+\eta')\Tr(s^\+s)_{\underline{1'}}
+\lambda_3[(\phi^\+s^\+)(s\phi')]_{\underline{1}}
+\lambda_4[(\eta^\+s^\+)(s\eta')]_{\underline{1}}\crn&&
+(\lambda_5\phi^\+\phi'
+\lambda_6\eta^\+\eta')\Tr(s'^\+s')_{\underline{1'}}
+\lambda_7[(\phi^\+s'^\+)(s'\phi')]_{\underline{1}}
+\lambda_8[(\eta^\+s'^\+)(s'\eta')]_{\underline{1}}\crn&&
+(\lambda_9\phi^\+\phi'
+\lambda_{10}\eta^\+\eta')\Tr(s^\+s')_{\underline{1'}}
+\lambda_{11}[(\phi^\+s^\+)(s'\phi')]_{\underline{1}}
+\lambda_{12}[(\eta^\+s^\+)(s'\eta')]_{\underline{1}}\crn&&
+(\lambda_{13}\phi^\+\phi'
+\lambda_{14}\eta^\+\eta')\Tr(s'^\+s)_{\underline{1'}}
+\lambda_{15}[(\phi^\+s'^\+)(s\phi')]_{\underline{1}}
+\lambda_{16}[(\eta^\+s'^\+)(s\eta')]_{\underline{1}}\crn&& +h.c.
\eea

To provide the Majorana masses for the neutrinos, the lepton
number must be broken. This can be achieved via the scalar
potential violating $U(1)_{\mathcal{L}}$, however the other
symmetries should be conserved. The $\mathcal{L}$ violating
potential is given as \bea \bar{V}&=&\overline{\mu}\chi\rho\eta'+
[\overline{\lambda}_1\phi^\+
\phi+\overline{\lambda}_2\phi'^\+\phi'+\overline{\lambda}_3\chi^\+
\chi+\overline{\lambda}_4\eta^\+\eta
+\overline{\lambda}_5\eta'^\+\eta'+\overline{\lambda}_6\rho^\+\rho+\overline{\lambda}_7\eta^\+
\chi+\overline{\lambda}_8\rho^\+\phi'\crn&&
+\overline{\lambda}_9\Tr(s^\+s)_{\underline{1}}+\overline{\lambda}_{10}\Tr(s'^\+s')_{\underline{1}}+\overline{\lambda}_{11}\Tr(s^\+s')_{\underline{1}}+\overline{\lambda}_{12}\Tr(s'^\+s)_{\underline{1}}](\eta^\+\chi)
+[\overline{\lambda}_{13}\phi^\+
\phi'+\overline{\lambda}_{14}\phi'^\+ \phi\crn&&
+\overline{\lambda}_{15}\eta^\+\eta'
+\overline{\lambda}_{16}\eta'^\+\eta+\overline{\lambda}_{17}\eta'^\+\chi+\overline{\lambda}_{18}\rho^\+\phi+\overline{\lambda}_{19}\Tr(s^\+s)_{\underline{1'}}+\overline{\lambda}_{20}\Tr(s'^\+s')_{\underline{1'}}+\overline{\lambda}_{21}\Tr(s^\+s')_{\underline{1'}}\crn&&+\overline{\lambda}_{22}\Tr(s'^\+s)_{\underline{1'}}
](\eta'^\+\chi)+[\overline{\lambda}_{23}\eta^\+\phi+\overline{\lambda}_{24}\eta'^\+\phi'+\overline{\lambda}_{25}\eta'^\+\rho](\phi^\+\chi)+[\overline{\lambda}_{26}
\eta^\+\phi'
+\overline{\lambda}_{27}\eta'^\+\phi+\overline{\lambda}_{28}\chi^\+\rho\crn&&
+\overline{\lambda}_{29}\eta^\+\rho](\phi'^\+\chi)
+[\overline{\lambda}_{30}\phi^\+\phi'+\overline{\lambda}_{31}\phi'^\+\phi+\overline{\lambda}_{32}\eta^\+\eta'
+\overline{\lambda}_{33}\eta'^\+\eta+\overline{\lambda}_{34}\Tr(s^\+s)_{\underline{1'}}+\overline{\lambda}_{35}\Tr(s'^\+s')_{\underline{1'}}\crn&&
+\overline{\lambda}_{36}\Tr(s^\+s')_{\underline{1'}}+\overline{\lambda}_{37}\Tr(s'^\+s)_{\underline{1'}}](\phi^\+\rho)
+[\overline{\lambda}_{38}\phi^\+\phi+\overline{\lambda}_{39}\phi'^\+\phi'+\overline{\la}_{40}\chi^\+\chi+\overline{\lambda}_{41}\eta^\+\eta+\overline{\lambda}_{42}\eta'^\+\eta'\crn&&
+\overline{\lambda}_{43}\Tr(s^\+s)_{\underline{1}}
+\overline{\lambda}_{44}\Tr(s'^\+s')_{\underline{1}}+\overline{\lambda}_{45}\Tr(s^\+s')_{\underline{1}}+\overline{\lambda}_{46}\Tr(s'^\+s)_{\underline{1}}](\phi'^\+\rho)
+[\overline{\lambda}_{47}\phi^\+\eta'+\overline{\lambda}_{48}\phi'^\+\eta](\chi^\+\rho)\crn&&
+[\overline{\lambda}_{49}\phi^\+\eta'+\overline{\lambda}_{50}\phi'^\+\eta](\eta^\+\rho)
+[\overline{\lambda}_{51}\phi^\+\eta+\overline{\lambda}_{52}\phi'^\+\eta'](\eta'^\+\rho)
+\overline{\lambda}_{53}[(\eta^\+s^\+)(s\chi)]_{\underline{1}}
+\overline{\lambda}_{54}[(\eta^\+s'^\+)(s'\chi)]_{\underline{1}}\crn&&
+\overline{\lambda}_{55}[(\eta^\+s^\+)(s'\chi)]_{\underline{1}}
+\overline{\lambda}_{56}[(\eta^\+s'^\+)(s\chi)]_{\underline{1}}
+\overline{\lambda}_{57}[(\eta'^\+s^\+)(s\chi)]_{\underline{1}}
+\overline{\lambda}_{58}[(\eta'^\+s'^\+)(s'\chi)]_{\underline{1}}\crn&&
+\overline{\lambda}_{59}[(\eta'^\+s^\+)(s'\chi)]_{\underline{1}}
+\overline{\lambda}_{60}[(\eta'^\+s'^\+)(s\chi)]_{\underline{1}}
+\overline{\lambda}_{61}[(\rho^\+s^\+)(s\phi)]_{\underline{1}}
+\overline{\lambda}_{62}[(\rho^\+s'^\+)(s'\phi)]_{\underline{1}}\crn&&
+\overline{\lambda}_{63}[(\rho^\+s^\+)(s'\phi)]_{\underline{1}}+\overline{\lambda}_{64}[(\rho^\+s'^\+)(s\phi)]_{\underline{1}}+\overline{\lambda}_{65}[(\rho^\+s^\+)(s\phi')]_{\underline{1}}+\overline{\lambda}_{66}[(\rho^\+s'^\+)(s'\phi')]_{\underline{1}}\crn&&
+\overline{\lambda}_{67}[(\rho^\+s^\+)(s'\phi')]_{\underline{1}}+\overline{\lambda}_{68}[(\rho^\+s'^\+)(s\phi')]_{\underline{1}}
+\overline{\lambda}_{69}\phi[(\phi
s^\+)s^\+]_{\underline{1}}+\overline{\lambda}_{70}\phi[(\phi
s'^\+)s'^\+]_{\underline{1}}+\overline{\lambda}_{71}\phi[(\phi
s^\+)s'^\+]_{\underline{1}}\crn&&
+\overline{\lambda}_{72}\phi[(\phi s'^\+)s^\+]_{\underline{1}}
+\overline{\lambda}_{73}\phi'[(\phi'
s^\+)s^\+]_{\underline{1'}}+\overline{\lambda}_{74}\phi'[(\phi'
s'^\+)s'^\+]_{\underline{1'}}+\overline{\lambda}_{75}\phi'[(\phi'
s^\+)s'^\+]_{\underline{1'}}+\overline{\lambda}_{76}\phi'[(\phi'
s'^\+)s^\+]_{\underline{1'}}\crn&&
+\overline{\lambda}_{77}\phi[(\phi'
s^\+)s^\+]_{\underline{1}}+\overline{\lambda}_{78}\phi[(\phi'
s'^\+)s'^\+]_{\underline{1}}+\overline{\lambda}_{79}\phi[(\phi'
s^\+)s'^\+]_{\underline{1}}+\overline{\lambda}_{80}\phi[(\phi'
s'^\+)s^\+]_{\underline{1}} +\overline{\lambda}_{81}\phi'[(\phi
s^\+)s^\+]_{\underline{1'}}\crn&&+\overline{\lambda}_{82}\phi'[(\phi
s'^\+)s'^\+]_{\underline{1'}}+\overline{\lambda}_{83}\phi'[(\phi
s^\+)s'^\+]_{\underline{1'}}+\overline{\lambda}_{84}\phi'[(\phi
s'^\+)s^\+]_{\underline{1'}}+h.c. \label{vbar} \eea

\subsection{The 3-3-1 model with right-handed neutrinos}
The general potential is given as that of the 3-3-1 model with
neutral fermions but all the terms relevant to the $\rho$ scalar
should be suppressed. In addition, $V_{\mathrm{tri-sext}}$
contains extra terms as follows: \bea && \lambda_{17}\phi[(\phi
s^\+)s^\+]_{\underline{1}}+\lambda_{18}\phi[(\phi
s'^\+)s'^\+]_{\underline{1}}+\lambda_{19}\phi[(\phi
s^\+)s'^\+]_{\underline{1}}+\lambda_{20}\phi[(\phi
s'^\+)s^\+]_{\underline{1}}\crn&& +\lambda_{21}\phi'[(\phi'
s^\+)s^\+]_{\underline{1'}}+\lambda_{22}\phi'[(\phi'
s'^\+)s'^\+]_{\underline{1'}}+\lambda_{23}\phi'[(\phi'
s^\+)s'^\+]_{\underline{1'}}+\lambda_{24}\phi'[(\phi'
s'^\+)s^\+]_{\underline{1'}}\crn&& +\lambda_{25}\phi[(\phi'
s^\+)s^\+]_{\underline{1}}+\lambda_{26}\phi[(\phi'
s'^\+)s'^\+]_{\underline{1}}+\lambda_{27}\phi[(\phi'
s^\+)s'^\+]_{\underline{1}}+\lambda_{28}\phi[(\phi'
s'^\+)s^\+]_{\underline{1}}\crn&& +\lambda_{29}\phi'[(\phi
s^\+)s^\+]_{\underline{1'}}+\lambda_{30}\phi'[(\phi
s'^\+)s'^\+]_{\underline{1'}}+\lambda_{31}\phi'[(\phi
s^\+)s'^\+]_{\underline{1'}}+\lambda_{32}\phi'[(\phi
s'^\+)s^\+]_{\underline{1'}}\crn&& +h.c.\label{m2} \eea It is
noted that in the 3-3-1 model with neutral fermions the similar
couplings appear, however, in the $\mathcal{L}$ violating
potential (\ref{vbar}). Therefore it will disappear in the one
mentioned below.

The $\mathcal{L}$ violating potential for this model is similar to
that of the 3-3-1 model with neutral fermions (\ref{vbar}),
however all the interactions therein that have appeared in
(\ref{m2}) must be removed.


\begin{thebibliography}{99}

\bibitem{pdg} K. Nakamura \emph{et al.} [Particle Data Group], J. Phys. G \textbf{37}, 075021 (2010).

\bibitem{seesaw1}
P. Minkowski, Phys. lett. B {\bf 67}, 421 (1977); M. Gell-Mann, P.
Ramond and R. Slansky, {\it Complex spinors and unified theories},
in {\it Supergravity}, edited by P. van Nieuwenhuizen and D. Z.
Freedman (North Holland, Amsterdam, 1979), p. 315; T. Yanagida, in
{\it Proceedings of the Workshop on the Unified Theory and the
Baryon Number in the Universe}, edited by O. Sawada and A.
Sugamoto (KEK, Tsukuba, Japan, 1979), p. 95; S. L. Glashow, {\it
The future of elementary particle physics}, in {\it Proceedings of
the 1979 Carg{\`e}se Summer Institute on Quarks and Leptons},
edited by M. L{\'e}vy et al. (Plenum Press, New York, 1980), pp.
687-713; R. N. Mohapatra and G. Senjanovi{\'c}, Phys. Rev. Lett.
{\bf 44}, 912 (1980).

\bibitem{leptog} M. Fukugita and T. Yanagida, Phys. Lett. B {\bf 174},
45 (1986).

\bibitem{lrm} J. C. Pati and A. Salam, Phys. Rev. D {\bf 10}, 275
(1974); R. N. Mohapatra and J. C. Pati, Phys. Rev. D {\bf 11},
566, 2558 (1975); G. Senjanovi{\'c} and R. N. Mohapatra, Phys.
Rev. D {\bf 12}, 1502 (1975).

\bibitem{so10} H. Georgi, in {\it Particles and Fields},
edited by C. E. Carlson (A.I.P., New York, 1975); H. Fritzsch and
P. Minkowski, Ann. Phys. {\bf 93}, 193 (1975).

\bibitem{331r} M. Singer, J. W. F. Valle and J. Schechter, Phys.
Rev. D {\bf 22}, 738 (1980); J. C. Montero, F. Pisano and V.
Pleitez, Phys. Rev. D {\bf 47}, 2918 (1993); R. Foot, H. N. Long
and Tuan A. Tran, Phys. Rev. D {\bf 50}, 34(R) (1994); H. N. Long,
Phys. Rev. D {\bf 53}, 437 (1996); {\bf 54}, 4691 (1996).

\bibitem{331m} F. Pisano and V. Pleitez, Phys. Rev.  D {\bf 46}, 410 (1992);
P. H. Frampton, Phys. Rev. Lett. {\bf 69}, 2889 (1992); R. Foot,
O. F. Hernandez, F. Pisano and V. Pleitez, Phys. Rev. D {\bf 47},
4158 (1993).

\bibitem{ecn331} W. A. Ponce, Y. Giraldo
and L. A. Sanchez, Phys. Rev. D {\bf 67}, 075001 (2003); P. V.
Dong, H. N. Long, D. T. Nhung and D. V. Soa, Phys. Rev. D {\bf
73}, 035004 (2006); P. V. Dong and H. N. Long, Adv. High Energy
Phys. {\bf 2008}, 739492 (2008).

\bibitem{anoma} A. Doff and F. Pisano, Mod. Phys. Lett. A {\bf 15}, 1471
(2000); P. V. Dong and H. N. Long, Int. J. Mod. Phys. A {\bf 21}
6677, (2006).

\bibitem{dlshA4} P. V. Dong, L. T. Hue, H. N. Long
and D. V. Soa, Phys. Rev. D {\bf 81}, 053004 (2010).

\bibitem{dlsvS4} P. V. Dong, H. N. Long, D. V. Soa, and V. V. Vien, Eur.
Phys. J. C \textbf{71}, 1544 (2011).

\bibitem{hps} P. F. Harrison, D. H. Perkins and W. G. Scott, Phys. Lett. B {\bf 530}, 167
(2002); Z. Z. Xing, Phys. Lett. B {\bf 533}, 85 (2002); X. G. He
and A. Zee, Phys. Lett. B {\bf 560}, 87 (2003); Phys. Rev. D {\bf
68}, 037302 (2003).

\bibitem{A4}
E. Ma and G. Rajasekaran, Phys. Rev. D {\bf 64}, 113012 (2001); K.
S. Babu, E. Ma and J. W. F. Valle, Phys. Lett. B \textbf{552}, 207
(2003); G. Altarelli and F. Feruglio, Nucl. Phys. B \textbf{720},
64 (2005); E. Ma, Phys. Rev. D {\bf 73}, 057304 (2006); X. G. He,
Y. Y. Keum and R. R. Volkas, JHEP {\bf 0604}, 039 (2006); S.
Morisi, M. Picariello, and E. Torrente-Lujan, Phys. Rev. D {\bf
75}, 075015 (2007); C. S. Lam, Phys. Lett. B {\bf 656}, 193
(2007); F. Bazzocchi, S. Kaneko and S. Morisi, JHEP \textbf{0803},
063 (2008); A. Blum, C. Hagedorn, and M. Lindner, Phys. Rev. D
{\bf 77}, 076004 (2008); F. Bazzochi, M. Frigerio, and S. Morisi,
Phys. Rev. D {\bf 78}, 116018 (2008); G. Altarelli, F. Feruglio
and C. Hagedorn, JHEP \textbf{0803}, 052 (2008); M. Hirsch, S.
Morisi and J. W. F. Valle, Phys. Rev. D {\bf 78}, 093007 (2008);
E. Ma, Phys. Lett. B {\bf 671}, 366 (2009); G. Altarelli and D.
Meloni, J. Phys. G {\bf 36}, 085005 (2009); Y. Lin, Nucl.\ Phys.\
B {\bf 813}, 91 (2009); Y. H. Ahn and C. S. Chen, Phys. Rev. D
{\bf 81}, 105013 (2010); J. Barry and W. Rodejohanny, Phys. Rev. D
{\bf 81}, 093002 (2010); 119901(E) (2010); G. J. Ding and D.
Meloni, Nucl. Phys. B {\bf 855}, 21 (2012).

\bibitem{S4}
 R. N. Mohapatra, M. K. Parida, G. Rajasekaran, Phys. Rev. D \textbf{69}, 053007 (2004);
 C. Hagedorn, M. Lindner, and R. N. Mohapatra, JHEP \textbf{0606}, 042 (2006);
 E. Ma, Phys. Lett. B \textbf{632}, 352 (2006);
 H. Zhang, Phys. Lett. B \textbf{655}, 132 (2007);
 Y. Koide, JHEP \textbf{0708}, 086 (2007);
 H. Ishimori, Y. Shimizu, and M. Tanimoto, Prog. Theor. Phys. \textbf{121}, 769 (2009);
F. Bazzocchi, L. Merlo, and S. Morisi, Nucl. Phys. B {\bf 816},
204 (2009); F. Bazzocchi and S. Morisi, Phys. Rev. D \textbf{80}
096005 (2009); G. Altarelli and F. Fergulio, Rev. Mod. Phys. {\bf
82}, 2701 (2010); G.J. Ding, Nucl. Phys. B \textbf{827}, 82
(2010); Y. H. Ahn, S. K. Kang, C. S. Kim, and T. P. Nguyen, Phys.
Rev. D {\bf 82}, 093005 (2010); Y. Daikoku and H. Okada,
arXiv:1008.0914 [hep-ph]; H. Ishimori, Y. Shimizu, M. Tanimoto,
and A. Watanabe, Phys. Rev. D {\bf 83} 033004 (2011); H. Ishimori
and M. Tanimoto, Prog. Theor. Phys. {\bf 125}, 653 (2011); R. Z.
Yang and H. Zhang, Phys. Lett. B {\bf 700} 316, (2011); S. Morisi
and E. Peinado, Phys. Lett. B {\bf 701}, 451 (2011); S. Morisi, K.
M. Patel, and E. Peinado, Phys. Rev. D {\bf 84}, 053002 (2011).

\bibitem{dlprd} P. V. Dong and H. N. Long, Phys. Rev. D {\bf 77}, 057302
(2008).

\bibitem{mal} K. Abe {\it et al.} [T2K Collaboration], Phys. Rev. Lett. {\bf 107}, 041801
(2011); P. Adamson {\it et al}. [MINOS Collaboration],
arXiv:1108.0015 [hep-ex].

\bibitem{vale} T. Schwetz, M. Tortola, and J. Valle, New J. Phys. {\bf
13}, 109401 (2011).

\bibitem{flivale} G. L. Fogli, E. Lisi, A. Marrone, A. Palazzo, and A. M.
Rotunno, Phys. Rev. Lett. {\bf 101}, 141801 (2008); M. C.
Gonzalez-Garcia, M. Maltoni, and J. Salvado, JHEP {\bf 1004}, 056
(2010); G. L. Fogli {\it et al.}, Phys. Rev. D {\bf 84} 053007
(2011); T. Schwetz, M. Tortola, and J. W. F. Valle, New J. Phys.
{\bf 13}, 063004 (2011).

\bibitem{kj} E. Ma, arXiv:hep-ph/0409075; H. Ishimori \textit{et. al.},
Prog. Theor. Phys. Suppl. \textbf{183}, 1 (2010).

\bibitem{s3model} See, for an incomplete list,
L. Wolfenstein, Phys. Rev. D {\bf 18}, 958 (1978);
S. Pakvasa and H. Sugawara, Phys. Lett. B {\bf 73}, 61 (1978);
{\bf 82}, 105 (1979); E. Durman and H. S. Tsao, Phys. Rev. D {\bf
20}, 1207 (1979); Y. Yamanaka, H. Sugawara, and S. Pakvasa, Phys.
Rev. D {\bf 25}, 1895 (1982); K. Kang, J. E. Kim, and P. Ko, Z.
Phys. C {\bf 72}, 671 (1996); H. Fritzsch and Z. Z. Xing, Phys.
Lett. B {\bf 372}, 265 (1996); K. Kang, S. K. Kang, J. E. Kim, and
P. Ko, Mod. Phys. Lett. A {\bf 12}, 1175 (1997); M. Fukugita, M.
Tanimoto, and T. Yanagida, Phys. Rev. D {\bf 57}, 4429 (1998); H.
Fritzsch and Z. Z. Xing, Phys. Lett. B {\bf 440}, 313 (1998); Y.
Koide, Phys. Rev. D {\bf 60}, 077301 (1999); H. Fritzsch and Z. Z.
Xing, Phys. Rev. D {\bf 61}, 073016 (2000); M. Tanimoto, Phys.
Lett. B {\bf 483}, 417 (2000); G. C. Branco and J. I.
Silva-Marcos, Phys. Lett. B {\bf 526}, 104 (2002); M. Fujii, K.
Hamaguchi and T. Yanagida, Phys. Rev. D {\bf 65}, 115012 (2002);
J. Kubo, A. Mondragon, M. Mondragon and E. Rodriguez-Jauregui,
Prog. Theor. Phys. {\bf 109}, 795 (2003); {\bf 114}, 287(E)
(2005); P. F. Harrison and W. G. Scott, Phys. Lett. B {\bf 557},
76 (2003); S.-L. Chen, M. Frigerio, and E. Ma, Phys. Rev. D {\bf
70}, 073008 (2004); {\bf 70}, 079905(E) (2004); H. Fritzsch and Z.
Z. Xing, Phys. Lett. B {\bf 598}, 237 (2004); F. Caravaglios and
S. Morisi, arXiv:hep-ph/0503234; W. Grimus and L. Lavoura, JHEP
{\bf 0508}, 013 (2005); R. N. Mohapatra, S. Nasri and H. B. Yu,
Phys. Lett. B {\bf 639}, 318 (2006); R. Jora, S. Nasri and J.
Schechter, Int. J. Mod. Phys. A {\bf 21}, 5875 (2006); J. E. Kim
and J.-C. Park, JHEP {\bf 0605}, 017 (2006); Y. Koide, Eur. Phys.
J. C {\bf 50}, 809 (2007); A. Mondragon, M. Mondragon, and E.
Peinado, Phys. Rev. D {\bf 76}, 076003 (2007); AIP Conf. Proc.
{\bf 1026}, 164 (2008); M. Picariello, Int. J. Mod. Phys. A {\bf
23}, 4435 (2008); C. Y. Chen and L. Wolfenstein, Phys. Rev. D {\bf
77}, 093009 (2008); R. Jora, J. Schechter and M. Naeem Shahid,
Phys. Rev. D {\bf 80}, 093007 (2009); {\bf 82}, 079902(E) (2010);
D. A. Dicus, S. F. Ge and W. W. Repko, Phys. Rev. D {\bf 82},
033005 (2010); Z. Z. Xing, D. Yang and S. Zhou, Phys. Lett. B {\bf
690}, 304 (2010); R. Jora, J. Schechter and M. N. Shahid, Phys.
Rev. D {\bf 82}, 053006 (2010); S. Dev, S. Gupta and R. R. Gautam,
Phys. Lett. B {\bf 702}, 28 (2011); D. Meloni, S. Morisi and E.
Peinado, J. Phys. G {\bf 38}, 015003 (2011); G. Bhattacharyya, P.
Leser and H. Pas, Phys. Rev. D {\bf 83}, 011701(R) (2011); T.
Kaneko and H. Sugawara, Phys. Lett. B {\bf 697}, 329 (2011); S.
Zhou, Phys. Lett. B {\bf 704}, 291 (2011).

\bibitem{clong} D. Chang and H. N. Long, Phys. Rev. {\bf D 73}, 053006 (2006).

\bibitem{lpa} E. Ma, Mod. Phys. Lett. A {\bf 25}, 2215 (2010).

\bibitem{malv} See, for example, E. Ma and U. Sarkar, Phys. Rev. Lett. {\bf 80}, 5716
(1998).

\bibitem{seesaw2} R. N. Mohapatra and G. Senjanovi{\'c}, Phys. Rev. D {\bf 23}, 165 (1981); G. Lazarides,
Q. Shafi and C. Wetterich, Nucl. Phys. B {\bf 181}, 287 (1981); J.
Schechter and J. W. Valle, Phys. Rev. D {\bf 25}, 774 (1982).

\bibitem{manf} E. Ma, Phys. Rev. Lett. {\bf 86}, 2502 (2001).

\bibitem{dls} P. V. Dong, H. N. Long and D. V. Soa, Phys. Rev. D {\bf 75}, 073006
(2007).







\end{thebibliography}
\end{document}